\documentclass[aps, prd, amsmath, floats, floatfix, twocolumn,
superscriptaddress, nofootinbib, showpacs]{revtex4}
\usepackage{graphicx}
\usepackage{epsfig}
\usepackage{color}
\usepackage{soul}
\usepackage{url}
\usepackage{times}
\usepackage{bm}         
\usepackage{times}

\newcommand{\beq}{\begin{equation}}
\newcommand{\eeq}{\end{equation}}
\newcommand{\beqn}{\begin{eqnarray}}
\newcommand{\eeqn}{\end{eqnarray}}

\newcommand{\tstrut}{\rule[-6pt]{0pt}{18pt}}
\newcommand{\tstruta}{\rule[-4pt]{0pt}{14pt}}
\newcommand{\tstrutb}{\rule{0pt}{2.7ex}}

\usepackage{color}

\newcommand{\Caltech}{\affiliation{Theoretical Astrophysics 350-17,
    California Institute of Technology, Pasadena, California 91125, USA}}
\newcommand{\Cornell}{\affiliation{Center for Radiophysics and Space
    Research, Cornell University, Ithaca, New York, 14853, USA}}
\newcommand{\WSU}{\affiliation{Department of Physics \& Astronomy,
	Washington State University, Pullman, Washington 99164, USA}}

\usepackage{graphicx}
\usepackage{dcolumn}
\usepackage{bm}
\usepackage{epsf}

\begin{document}

\title{Black hole-neutron star mergers: effects of the orientation of the black
hole spin}



\author{Francois Foucart} \Cornell %
\author{Matthew D. Duez} \Cornell\WSU %
\author{Lawrence E. Kidder} \Cornell %
\author{Saul A. Teukolsky} \Cornell\Caltech %

\begin{abstract}
The spin of black holes in black hole-neutron star (BHNS) binaries can have a
strong influence on the merger dynamics and the postmerger state; a wide
variety of spin magnitudes and orientations are expected to occur in
nature.  In this paper, we report the first simulations in full general
relativity of BHNS mergers with misaligned black hole spin.  We vary
the spin magnitude from $a_{\rm BH}/M_{\rm BH}=0$ to 
$a_{\rm BH}/M_{\rm BH}=0.9$ for aligned cases, and we vary the
misalignment angle from 0 to $80^\circ$ for $a_{\rm BH}/M_{\rm BH}=0.5$.
We restrict our study
to 3:1 mass ratio systems and use a simple $\Gamma$-law equation of state.  We
find that
the misalignment angle has a strong effect on the mass of the postmerger
accretion disk, but only for angles greater than $\approx 40^\circ$. 
Although the disk mass varies significantly with spin magnitude and
misalignment angle, we find that all disks have very similar lifetimes
$\approx 100$ms.  Their thermal and rotational profiles are also very
similar.  For a misaligned merger, the disk is tilted with respect to the
final black hole's spin axis.  This will cause the disk to precess, but on a
time scale longer than the accretion time.  In all cases, we find promising
setups for gamma-ray burst production:  the disks are
hot, thick, and hyperaccreting, and a baryon-clear region exists above the
black hole.

\end{abstract}


\maketitle

\section{Introduction}
\label{intro}
Black hole-neutron star (BHNS) binary mergers present
a remarkable opportunity to study strongly-curved spacetime and
supernuclear-density matter in the most extreme, dynamical conditions. 
BHNS binaries in compact orbits emit strong gravitational waves, and they
are expected to be one of the main sources for Advanced LIGO and
VIRGO~\cite{Ligo,Virgo}.  Current estimates for the event rates of binary
mergers coming from population synthesis models predict that Advanced LIGO
will see about $10$ BHNS/yr, although uncertainties in the models allow
for a large range of potential event rates, $\sim 0.2$ -- 300
BHNS/yr~\cite{Ligo:2010}.  These gravitational waves contain, in
principle, a wealth of information on their source, such as the mass and
spin of the black hole (BH) and the mass and radius of the neutron star (NS). 
Inferred properties of
the NS could be used to constrain the NS equation of state.  Information
in the waves can only be extracted, however, by comparison with accurate
numerically-generated predictions that provide the expected waveform for
each possible BHNS system.

Mergers of BHNS binaries have also been proposed as potential progenitors
of short-hard gamma-ray bursts (SGRB)~\cite{1991AcA....41..257P}. The
origin of SGRBs is not yet known, although it is certain that the engines
are compact and located at cosmological distances, and there is evidence
(such as their presence in nonstar forming regions) to support a mechanism
different from that associated with long-soft GRBs, namely stellar core
collapse.  For BHNS mergers, the generation of a SGRB is possible only if
the remnant black hole is surrounded by a massive, 
hot, thick accretion disk. Also, to obtain relativistic jets and a 
beamed outflow, a region mostly devoid of any matter is necessary 
(see e.g.~\cite{2007NJPh....9...17L} and references therein).  Only
numerical simulations in full general relativity with realistic microphysics
can determine if these conditions are likely to be obtained.

Whether a disk forms or not will depend on the premerger
characteristics of the binary, especially the BH mass, the NS radius,
and the BH spin.  Current estimates from population synthesis
models suggest that most systems are likely to be formed with a
black hole of $\sim 10M_\odot$.  Relativistic simulations to date have
considered cases of relatively low mass black holes ($\sim 2-7M_\odot$
)~\cite{Shibata:2006ks,Shibata:2006bs,Shibata:2007zm,Shibata:2009cn,
Etienne:2007jg,Duez:2008rb,Etienne:2008re,Duez:2010a}, 
for which the NS
is expected to disrupt outside the innermost stable circular orbit (ISCO),
making disk formation more likely.  These simulations have found cases
of massive disk formation, with $M_{\rm BH}\sim 3-4M_\odot$ resulting in
the largest disks~\cite{Etienne:2008re}.  The NS radius is the parameter
related to the equation of state that has the largest effect on the waveform
and post-merger disk~\cite{Duez:2010a}, with larger radii resulting in
larger disks~\cite{Shibata:2007zm,Duez:2010a}.

The spin of the black hole can have a strong influence on the merger. 
The ISCO is smaller for prograde orbits around a spinning BH than for
orbits around a nonspinning hole.  Because disk formation is expected to
be more likely if NS tidal disruption occurs outside the ISCO than if it
occurs inside, BH spin can facilitate disk formation.  With high BH spin,
it is even plausible that BHNS binaries with the most likely mass ratios
($\sim$7:1) give rise to substantial disks~\cite{ferrariBHNS:09}. 
The magnitude of the BH spin is largely unconstrained by population synthesis
models, as it comes mostly from the spin acquired during formation
of the black hole
in a core-collapse event~\cite{2008ApJ...682..474B}.  The effect of
aligned and antialigned spins was investigated in full general relativity
for the 3:1 mass-ratio case by Etienne~{\it et al.}~\cite{Etienne:2008re}. 
They found that a large aligned spin (and correspondingly small ISCO)
leads to a much more massive post-merger disk. 
For example,
for $a_{\rm BH}/M_{\rm BH}=0.75$ and $M_{\rm BH}\sim 4M_\odot$, a disk
of $M_{\rm disk}\sim0.2M_\odot$ can be obtained. For BHNS binaries with
massive black holes  ($M_{\rm BH}\sim 10M_\odot$), forming a disk may
in fact only be possible if the hole is spinning.

There is no reason to expect the black hole spin to be aligned with the
orbital angular momentum.  Population synthesis models predict a relatively
wide distribution of orientations,
with about half of the binaries having a misalignment
between the BH spin and the orbital angular momentum of less than
$45^\circ$ when the initial BH spin is
$a_{\rm BH}/M_{\rm BH}=0.5$~\cite{2008ApJ...682..474B}.  
Misalignment can reduce or reverse the
BH spin effects described above. This can be understood
by considering prograde orbits of test 
particles with small radial velocity, which become 
unstable farther away from the BH for inclined 
orbits than for equatorial orbits of the same angular momentum. 
Misalignment will also produce
qualitatively new effects, including the precession of the premerger orbital
plane and BH spin.  The influence of misalignment has been studied for 10:1
mass-ratio binaries in the approximation that the spacetime is assumed to be
Kerr.  Rantsiou~{\it et al.}~\cite{2008ApJ...680.1326R} showed that, in
this approximation, disks can be formed only at relatively low inclinations
and only for near extremal black holes.  Ultimately, though, simulations in
full general relativity are needed to accurately model such systems.

In this paper, we report on fully relativistic studies of misaligned
BHNS binaries.  We limit ourselves to small mass systems
($M_{\rm BH}\sim 4.2M_\odot$) and a simplified equation of state, but
we consider a significant range of black hole spin magnitudes and
orientations.  We confirm the results of
Etienne~{\it et al}~\cite{Etienne:2008re} regarding the effects of an aligned
BH spin. For misaligned spins,
we find that the misalignment angle can have a strong effect on
the post-merger disk mass, but only for angles greater than around $40^\circ$. 
Although the disk mass varies greatly with BH spin, most other disk
properties are very similar, including the accretion time scale, the
location of the maximum density, the average temperature, and the
entropy and angular momentum profiles.  The disks are all thick, each with
a height $H$ to radius $r$ ratio $H/r\approx 0.2$, nearly independent of $r$. 
Crucially, they all have a baryon-clear region above
and below the BH.  The disks are misaligned with
the final BH spin by $\leq15^\circ$.  They do precess about the BH
spin axis, but without reaching a fixed precession rate. Indeed, the 
steady-state precession time scale is expected to be significantly longer than
the accretion time scale.  

This paper is organized as follows. In Sec.~\ref{methods}, we discuss
the method used to construct the very general BHNS initial data we use. We also
discuss in detail the improvements to our evolution code that have increased
the accuracy by an order of magnitude over the results presented in
Duez {\it et al.}~\cite{Duez:2008rb}.  
We then present our run diagnostics
in Sec.~\ref{diag}.  The different cases to be evolved are described in
Sec.~\ref{cases}.  We then present the results of the simulations in
Secs.~\ref{nonspinning} and~\ref{results}.  Finally, we draw conclusions
in Sec.~\ref{conclusions}.

\section{Numerical methods}
\label{methods}

\subsection{Initial data}
\label{init}

For numerical evolutions of Einstein's equations, we decompose the spacetime
under study into a foliation of spacelike hypersurfaces parametrized 
by the coordinate $t$.
Einstein's equations can be written in the form of hyperbolic evolution
equations plus a set of constraints that have to be satisfied on each 
$t={\rm constant}$ slice.  Our initial 
data at $t=0$ must be chosen such that it satisfies
these constraints. We construct initial data using the Extended Conformal
Thin Sandwich formalism (XCTS)~\cite{York:1999,Pfeiffer:2003}. 
If we write the spacetime metric as
\beqn
ds^2 &=& g_{\mu \nu} dx^{\mu} dx^{\nu} \nonumber \\
&=& -\alpha^2 dt^2 + \psi^4 \tilde{\gamma}_{ij} (dx^i + \beta^i dt)(dx^j + 
\beta^j dt),
\eeqn
the initial data to be determined include the lapse $\alpha$, the shift
vector $\beta^i$, the conformal factor $\psi$, the conformal 3-metric 
$\tilde{\gamma}_{ij}$ and the extrinsic curvature
$K_{\mu \nu}=-\tfrac{1}{2} {\cal L}_n g_{\mu \nu}$ (where ${\cal L}_n$ is the 
Lie derivative along the normal ${\bm n}$ to the $t=0$ slice).  
The constraints can be expressed as a set of 5 coupled elliptic 
equations for the lapse, shift, and conformal factor~\cite{Pfeiffer:2003}. 
The physical
properties of the system are then determined by the choice of the remaining
free parameters: the trace of the extrinsic curvature 
$K=g_{\mu \nu} K^{\mu \nu}$, the conformal metric $\tilde{\gamma}_{ij}$, 
their time derivatives $\partial_t K$ and $\partial_t \tilde{\gamma}_{ij}$, and
the matter stress-energy tensor $T_{\mu\nu}^{\rm matter}$.

The system of elliptic equations is solved using the spectral elliptic solver 
{\sc spells} developed by the Cornell-Caltech 
collaboration~\cite{Pfeiffer:2003a}, and initially used to construct 
initial data for binary black hole systems by Pfeiffer 
{\it et al.}~\cite{Pfeiffer:2002,Pfeiffer:2003}. A detailed presentation of 
the methods used for the construction of BHNS initial data was given in 
Foucart {\it et al.}~\cite{Foucart:2008a}. 
Here, we limit ourselves to a brief summary plus
a description of the changes made to accommodate the possibility of 
arbitrary spin orientation.

As the system is expected to be initially in a
quasiequilibrium state, with the binary in a low-eccentricity circular orbit
of slowly decreasing radius, we work in a frame comoving with the binary
and set the time derivatives to zero: $\partial_tK=0$ and 
$\partial_t\tilde{\gamma}_{ij}=0$. As for 
$\tilde{\gamma}_{ij}$ and $K$, we make a choice inspired by the results of 
Lovelace {\it et al.}~\cite{Lovelace:2008a} for binary black hole systems. 
Close to the BH, 
the metric matches its Kerr-Schild values for a BH of the desired mass
and spin, while away from the BH, the conformal metric is flat and $K=0$.
The transition between these two regions is done by using the following
prescription:
\beqn
\tilde \gamma_{ij}&=&\delta_{ij}+[\gamma_{ij}^{KS}(a_{\rm BH},
{\bf v}_{\rm BH})-\delta_{ij}] e^{-\lambda (r_{\rm BH}/w)^4},\\
K&=&K^{KS}(a_{\rm BH},{\bf v}_{\rm BH}) e^{-\lambda (r_{\rm BH}/w)^4},\\
\lambda &=& \frac{r_{\rm BH}-r_{\rm AH}}{r_{\rm NS}/q - r_{\rm BH}},\\
{\bf v}_{\rm BH}&=&{\bf \Omega}^{\rm rot} \times {\bf c}_{\rm BH},
\eeqn
where the $KS$ subscript refers to the Kerr-Schild values, $r_{\rm BH}$ 
($r_{\rm NS}$) is the coordinate distance to the center of the BH (NS), 
$r_{\rm AH}$ the coordinate radius of the BH apparent horizon,
$a_{\rm BH}/M_{\rm BH}$ is the dimensionless spin of the BH, 
$q\sim M_{\rm NS}/M_{\rm BH}$ a constant of the order of the mass ratio,
${\bf c}_{\rm BH}$ the coordinate location of the BH center
with respect to the center of mass of the system, and $w$ is some freely 
specifiable width, chosen so that the
metric is nearly flat at the location of the NS.
The parameter $\lambda$, which is designed to impose a flat
background at the location of the NS, is set to $\infty$ for 
$r_{\rm NS}<qr_{\rm BH}$.

Boundary conditions are imposed at infinity and on the apparent horizon of
the BH (since
the inside of the BH is excised from our computational domain). 
The boundary conditions at infinity
are chosen so that the metric is asymptotically flat, while the 
inner boundary conditions
follow the prescriptions of Cook and Pfeiffer~\cite{Cook:2004}, which 
make the inner boundary an apparent horizon in quasiequilibrium. 
There is some freedom in these boundary conditions: on the apparent horizon, 
the conformal lapse is not fixed (we set it to the value of an isolated 
Kerr BH), and the shift is determined only up to a rotation term 
$\beta'^i = \beta^i + \epsilon^{ijk} \Omega^{\rm BH}_j x_k$. 
The value of ${\bm \Omega}^{\rm BH}$ determines the spin of the BH, but the 
exact 
relation between ${\bm \Omega}^{\rm BH}$ and the spin is unknown {\it a priori};
to get the desired BH spin, we have to solve iteratively for 
${\bm \Omega}^{\rm BH}$. On the outer
boundary, the shift can be written as
\beq
{\bm \beta}={\bm \Omega}^{\rm rot} \times {\bf r} + {\dot a}_0 {\bf r} 
+ {\bm v}_{\rm boost}
\eeq
where ${\bm \Omega}^{\rm rot}$ allows for a global rotation of the coordinates,
$\dot{a_0}$ for a radial infall with velocity ${\bf v}={\dot a}_0 {\bf r}$, and
${\bm v}_{\rm boost}$ for a boost. As an initial guess for the
orbit of the binary, we can set the radial velocity at $t=0$ to 0
($\dot{a_0}=0$). This assumption, as well as the quasiequilibirum formalism,
clearly neglects the evolution of the orbit
over time through the radial infall of the binary and the precession
of the orbital plane. Both effects are, however, acting over relatively
long time scales: over its first orbit, even the binary with the most inclined 
spin considered here (s.5i80 in the later sections) goes through less than $10\%$
of a full precession period of the BH spin while the coordinate separation
between the compact objects is reduced by about $20\%$. One known effect
of the quasi-circular approximation is that the binary will have a nonzero
eccentricity. The eccentricity can be decreased by modifying the initial values of 
$\dot{a_0}$ and ${\bm \Omega}^{\rm rot}$~\cite{Pfeiffer:2007a}
(see also~\cite{Foucart:2008a} for an application of that method to BHNS
binaries) as long as the initial eccentricity and orbital phase can
be accurately measured. Here, we only apply this technique when the spin
of the BH is aligned with the orbital angular momentum of the binary. For
precessing binaries, significantly reducing the eccentricity would require
a larger initial separation for which the effects of eccentricity, precession
and radial infall can be properly disantangled.

In the presence of matter, additional choices are required. We assume that the 
fluid is in hydrostatic equilibrium in the comoving frame, and require that 
its 3-velocity is irrotational. The first condition gives 
an algebraic relation between the enthalpy $h$ of the fluid, its 3-velocity 
$v_i$, and the metric $g_{\mu \nu}$, while the second leads to another 
elliptic equation determining the velocity field. These 
equations are coupled to the constraints: the whole system can only be solved 
through an iterative method. For a BH with a spin aligned with the total
angular momentum of the system, that method is described in 
Foucart {\it et al.}~\cite{Foucart:2008a}: we solve for the metric 
using {\sc spells}, then 
determine the new value of the enthalpy $h$, as well as the orbital angular 
velocity ${\bm \Omega}^{\rm rot}$ (chosen so that the binary is in 
quasicircular orbit), the 
position of the BH in the equatorial plane (so that the total linear momentum 
$P_{\rm ADM}$ vanishes), and the free parameter $\Omega_{\rm BH}^z$ (to drive 
the spin of the BH to its desired value). Finally, we solve for the velocity 
field through the elliptic equation imposing an irrotational configuration, 
and go back to the first step. 

\begin{figure}
\includegraphics[width=8cm]{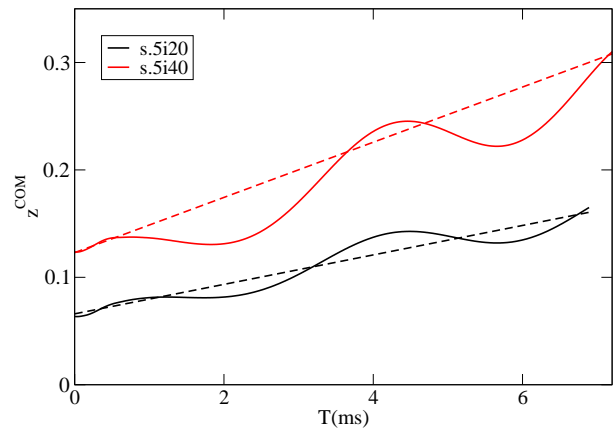}
\caption{
  Evolution of the position of the center of mass along the z-axis during 
inspiral (solid line), compared to the motion expected from the boost 
${\bm v}_{\infty}^z$ given in the initial data (dotted line).
}
\label{fig:PzID}
\end{figure}

In order to construct initial data for BHs with 
a spin that is not aligned with the orbital angular momentum of the binary, a 
few changes are necessary. First, we do not assume that ${\bm \Omega}^{\rm BH}$
is aligned with the orbital angular momentum. Instead, all 3 components of 
${\bm \Omega}^{\rm BH}$ are solved for. We also abandon the assumption of 
equatorial symmetry, and control the position of the BH along the $z$-axis of 
the orbital angular momentum by requiring that the NS is initially moving in 
the $xy$-plane, with its center in the $z=0$ plane (the $z$ coordinate of the 
location of the BH is chosen so that the condition 
$\hat{e}_z {\bf .} \nabla h = 0$ is satisfied at the center of the NS). 
Finally, to guarantee that $P_{\rm ADM}^z=0$ we impart a boost to the whole 
system through the boundary 
condition at infinity: ${\bm v}_{\rm boost}^z = v_{\infty}^z$. The center of 
mass is then expected to have a global motion during inspiral corresponding to 
that boost, and we check in Fig.~\ref{fig:PzID} that this is indeed the case.
By adding these conditions to the iterative procedure used to generate
BHNS initial data, we are able to obtain high-precision initial 
configurations for arbitrary values of the orientation of the BH spin.

\subsection{Evolution}
\label{evol}

The simulations presented here use the SpEC code developed 
by the Cornell-Caltech-CITA Collaboration~\cite{SpEC}. To evolve BHNS systems,
the two-grid method described in Duez {\it et al.}~\cite{Duez:2008rb}
is used. Einstein's equations are evolved on a pseudospectral 
grid, using the first-order generalized harmonic
formulation~\cite{Lindblom:2006}, while 
the hydrodynamical equations are solved on a separate finite difference grid
called the ``fluid grid''.  The hydrodynamic equations are written in
conservative form
\beq
\partial_t {\bf U} + \nabla {\bf F}({\bf U}) = {\bf S}({\bf U}).
\eeq
To compute the flux ${\bf F}$ on the faces of each finite difference cell,
we use the third-order shock capturing PPM reconstruction 
method~\cite{1984JCoPh..54..174C}.
More details on the numerical methods used can be found in 
Duez {\it et al.}~\cite{Duez:2008rb}. However, since the publication 
of \cite{Duez:2008rb},
several important improvements have been made to the code, which are
described in the following subsections.  

\subsubsection{Dynamic regridding}
\label{regridder}
To accurately evolve a BHNS binary while determining its gravitational wave 
emission, simulations have to resolve events occurring at very different scales.
When the neutron star is disrupted and a disk forms, we expect shocks in the
disk, and steep density and temperature gradients close to the BH.   
But the disruption of the star also leads to the creation of a long tidal tail 
which can initially contain up to $5-10\%$ of the initial mass of the 
star and expand hundreds of kilometers away from the center of the 
BH~\cite{Duez:2010a}. Clearly, both the sharp, small-scale features around the 
black hole and the large-scale tidal tail should be properly resolved if we
want to follow the formation of an accretion disk. Furthermore, to extract 
gravitational waves accurately, the evolution of the gravitational fields 
should extend to the wave zone, in regions where no matter at all is present.

Because we use different grids to evolve the metric and the fluid variables,
the spectral grid on which we solve the generalized harmonic equations can
be extended into the wave zone while the fluid grid used for 
the relativistic hydrodynamical equations only covers the region where  
matter is present. Our earlier simulations~\cite{Duez:2008rb} were limited to 
nonspinning black holes. In that case,
most of the matter was rapidly accreted onto 
the hole and the tidal tails and accretion disks were small enough that 
manually expanding the fluid grid at a few chosen timesteps 
allowed us to resolve the evolution of the fluid at a reasonable computational 
cost. For spinning black holes, 
this is no longer the case: cost-efficient evolutions require a grid with 
points concentrated in the high-density regions close to the BH, 
and coarser resolution in the tail. Furthermore, as the evolution of the fluid 
is highly dynamical, interrupting the simulation whenever the finite 
difference grid is no longer adapted to the fluid configuration becomes 
impractical. One solution would be to use an adaptive mesh 
refinement scheme, similar to the codes used by Yamamoto et 
al.~\cite{Yamamoto:2008js} and
Etienne et al.~\cite{Etienne:2008re}. In our code, we choose instead to 
use a map between the fluid grid and the pseudospectral grid that
concentrates grid points in the region close to the black hole and 
automatically follows the evolution of the fluid.

To do this, we measure the outflow of matter across surfaces close to but
inside of the fluid grid boundaries. As soon as the outflow across one of
these surfaces 
crosses a given threshold (chosen so that the amount of matter leaving the 
grid over the whole simulation is negligible compared to the final mass of the 
accretion disk), the grid expands. The opposite is done on fixed surfaces 
farther away from the grid boundaries to force the grid to contract whenever 
the fluid is moving away from a boundary. The map itself is the combination of:
\begin{enumerate}
\renewcommand{\labelenumi}{(\roman{enumi})}
\item A translation of the center of the grid, to follow the general motion of 
the fluid
\item A linear scaling of each coordinate axis, to adapt to its expansions and 
contractions
\item A radial map smoothly transitioning from a high resolution region close 
to the black hole to a lower resolution region far away from it. The exact 
form of the map is
\beq
r'=\begin{cases}
r, & r<r_A\\
f(r)-f(r_A)+r_A, & r_A<r<r_B\\
\lambda (r-r_B) + f(r_B)-f(r_A)+r_A, & r>r_B
\end{cases}
\eeq
\beq
f(r) = r (ar^3+br^2+cr+d), 
\eeq
where the coefficients $(a,b,c,d)$ are chosen so that the map is $C^2$ at $r_A$
and $r_B$ and $\lambda$ is chosen
so that the grid is of the desired size. The radii 
$r_A$ and $r_B$ are fixed for the whole evolution and will determine
respectively the minimum resolution in the neighborhood of the black hole and 
the characteristic lengthscale of the transition between the high and low 
resolution regions.   
\end{enumerate} 

\subsubsection{Excision}
\label{exision}
We find that our hydrodynamics code is more stable near the
excision zone if we switch from PPM to the more diffusive MC
reconstruction~\cite{MCreconstruction}
in the vicinity of the excision zone.  Therefore, we replace the
face values determined by PPM reconstruction, $u_{R,L}{}^{PPM}$ with
a weighted average:
\begin{equation}
  u_{R,L} = fu_{R,L}{}^{MC} + (1-f)u_{R,L}{}^{PPM}\ ,
\end{equation}
where $f=1$ for $r<r_1 \sim 2r_{\rm ex}$ and $f=e^{-[(r-r_1)/r_1]^2}$ 
for $r>r_1$.

The MC face-value computation must be altered when its regular stencil
would extend into the excised region, and doing this properly turns out to
be important for stability.  Consider a one-dimensional problem with
grid points $x_n = n\Delta x$.  Then the face-value reconstruction of
the function $u_i$ from the left $u_{L,i-1/2}=u_{i-1/2-\epsilon}$ and
from the right $u_{R,i-1/2}=u_{i-1/2+\epsilon}$ must be adjusted as follows.
\newline
(i) If $x_i$ is in the excision zone, but $x_{i-1}$ is outside, \\
set $u_{L,i+1/2} = u_{i-1}$ and $u_{R,i+1/2} = u_{i-1}$ \\
(ii) If $x_i$ is in the excision zone, but $x_{i+1}$ is outside, \\
set $u_{L,i-1/2} = u_{i+1}$ and $u_{R,i-1/2} = u_{i+1}$ \\
(iii) If $x_i$ is outside the excision zone, but $x_{i-1}$ is inside, \\
set $u_{L,i-1/2} = u_{R,i-1/2}$ \\
(iv) If $x_i$ is outside the excision zone, but $x_{i+1}$ is inside, \\
set $u_{R,i+1/2} = u_{L,i+1/2}$

We also observed that the stability of our code close to the excision surface was
strongly affected by the details of the interpolation method chosen for the communication 
from finite difference to spectral grid in that region. Previously, the interpolation stencil was
shifted away from the excision boundary until the entire stencil was out of the
excision zone. This could lead to unstable evolutions or large interpolation
errors if the excision region happened to be located close to the boundary between
two subdomains of the finite difference grid, and acceptable stencils could only
be found far from the point we were interpolating to --- or could not be found
at all (to limit MPI communications, the stencil has to be entirely contained 
in one subdomain). Currently, we limit the displacement of the stencil to a
maximum of 3 grid point separations. If there is no good stencil within that
distance, we decrease the order of the interpolation, and keep doing so
until an acceptable stencil is found. 

Another interpolation method would be to forbid any displacement of the stencil,
and immediately drop to lower order as soon as part of the stencil lies within the
excision zone. Both algorithms are equally robust, but when tested on an actual BHNS
merger the first appeared to perform better at maintaining a smooth solution and low
constraint violations on the excision surface. Accordingly, we chose it as our
standard interpolation method and used it for all simulations presented in
this paper.

\subsubsection{Coordinate evolution}
\label{coords}

In the generalized harmonic formulation, the evolution of the 
inertial coordinates $x_a$ is given by the inhomogeneous wave equation
\beq
\nabla^b\nabla_b x_a = H_a,
\eeq
where $\nabla_b$ is the covariant derivative along $x_b$.
The evolution of the function $H_a(x_b)$ can be freely specified, but
its value on the initial slice $t=0$ is determined by the initial data (the
lapse and shift at $t=0$ fix the initial evolution of the gauge). While the
binary spirals in, we choose $\partial_t H_a(t,\tilde{x}_i)=0$ 
in the coordinate frame $\tilde{x}_i$ comoving with the system. 
In our previous paper~\cite{Duez:2008rb}, 
we changed the gauge evolution during the merger 
phase by damping $H_a$ exponentially in the comoving frame:
\begin{equation}
  H_a(t,\tilde{x}_i) = e^{-(t-t_d)/\tau}H_a(t_d,\tilde{x}_i)\ ,
\end{equation}
where $t_d$ is  the disruption time---the time at which we begin damping--- 
and $\tau$ is a damping time scale of order 10$M$ ($M$ being the 
total mass of the system). 
Further experimentation has shown that it is better not to change $H_a$
near the excision zone. In our current simulations, we set
\begin{eqnarray}
\frac{H_a(t,\tilde{x}_i)}{H_a(t_d,\tilde{x}_i)} &=& \left\{ Q(\tilde{r}) + [1-Q(\tilde{r})]e^{-(t-t_d)/\tau} \right\} \\
Q(\tilde{r}) &=& e^{-(\tilde{r}/\tilde{r}_{\rm ex})^2+1}
\end{eqnarray}
during the merger phase, where $\tilde{r}$ is the distance to
the center of the black hole in the comoving frame, 
and $\tilde{r}_{\rm ex}$ is the excision radius.

\subsubsection{Evolutions with fixed metric}
\label{Cowling}

During the merger of a BHNS binary, both the spacetime metric and the fluid
configuration are highly dynamical. Einstein's equations have to be solved
together with the conservative hydrodynamics equations, and the evolution of
that coupled system is computationally intensive. However, a few milliseconds
after merger, the BH remnant settles into a quasistationary state as it 
accretes slowly from the surrounding accretion disk. Then, the evolution of the
metric does not have a strong influence on the behavior of the
system. In numerical simulations, we can thus extract some information
on the long-term behavior of the final black hole-accretion disk system by
neglecting the evolution of the metric and only evolving the fluid
(c.f.~\cite{2008PhRvD..77d4001S}). Using this approximation, our code runs
about 4 times faster.

\begin{figure}
\includegraphics[width=8cm]{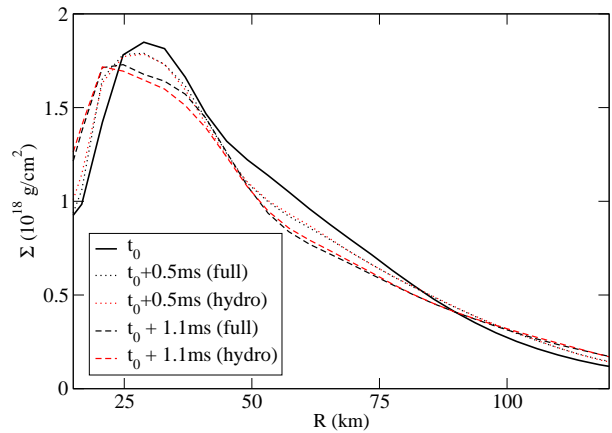}
\caption{
  Average surface density of the fluid at a given distance from the
  center of mass of the system, for the simulation s.5i20 described in 
  Sec.~\ref{cases}. The surface density is plotted at time 
  $t_0=t_{\rm merger}+10.3{\rm ms}$, when we begin to evolve the system using
  the fixed-metric approximation, as well as 0.5ms and 1.1ms later. We see
  that the profile is very similar for both evolution methods, even though
  the disk itself is not in a stationary configuration.
}
\label{fig:CowlingTest}
\end{figure}

To test the limitations of this method, we evolve the coupled system 
$\sim 1{\rm ms}$ past the time at which we begin the approximate, fluid-only 
evolution. We look for differences in the accretion rate or the characteristics
of the disk (temperature, density, inclination) between the two methods used. 
As long as
we wait for the properties of the black hole to settle down before switching
to the approximate evolution scheme, the two methods show extremely good
agreement --- except for the highest spin configuration, which 
leads to a massive disk that cannot be evolved accurately by fixing the 
background metric. In Fig.~\ref{fig:CowlingTest} we show the evolution 
of the density profile of a disk using both the full GR evolution and the
fixed-metric approximation. The evolution of the accretion disk is mostly
unaffected by the change of evolution scheme.

\section{Diagnostics}
\label{diag}

An indispensable test of numerical accuracy is convergence with grid
resolution.  We have evolved most of the cases discussed below at three
resolutions.  We call these Res1, Res2, and Res3; they correspond to
$100^3$, $120^3$, and $140^3$ gridpoints, respectively, on the fluid grid
and to $69^3$, $82^3$, and $95^3$ collocation points on the pseudospectral
grid.

The black hole is described by its irreducible mass $M_{\rm irr}$, its
spin $\bf{S}_{\rm BH}$, and its Christodoulou mass
$M_{\rm BH} = \sqrt{M_{\rm irr}^2 + S_{\rm BH}{}^2/(4M_{\rm irr}^2)}$. 
The spin $\bf{S}_{\rm BH}$ is computed using the approximate Killing vector
method~\cite{Lovelace:2008a}.

To monitor the nuclear matter, we first measure the 
baryonic mass $M_{\rm b}$ on the grid as a function of time.  Initially,
this will be the baryonic
mass of the neutron star.  After the tidal destruction of the star, it will
be the sum of the
accretion disk and the tidal tail masses.  At late times (more than 10ms after
merger), it will be the baryonic mass of the disk.  The accretion time scale
$\tau_{\rm disk}$ is $M_{\rm b}/(dM_{\rm b}/dt)$.

We also analyze the heating in the disk.  This is done through both an
entropy and a temperature variable.  When evolving with microphysical
equations of state, the temperature and entropy are provided directly. 
For this study, we use a $\Gamma$-law equation of state, so we need a way
to estimate the entropy and temperature.  The entropy measure $s$ we define as
$s=\log(\kappa/\kappa_i)$, where $\kappa$ is the polytropic constant obtained
from the relation between the pressure and the baryon density 
($P=\kappa \rho_0^\Gamma$), and
$\kappa_i$ is its initial, cold value.  To estimate the physical temperature,
$T$, we assume that the thermal contribution to the specific internal energy
$\epsilon_{\rm th} = \epsilon - \epsilon(T=0)$ is given by a sum of ideal
gas and blackbody components:
\beq
\epsilon_{\rm th} = {3 kT\over 2m_n} + f {a T^4\over \rho}\ ,
\eeq
where $m_n$ is the nucleon mass, and the factor $f$ reflects the number
of relativistic particles, and is itself a function of $T$. 
(See~\cite{Shibata:2007zm,Etienne:2008re}, who also make this assumption.) 
For the most part, we will report density-averaged values of $s$ and $T$. 
For example, the density-averaged entropy is
\beq
\langle s\rangle = \frac{\int \rho({\bf r}) s({\bf r}) dV}
{\int \rho({\bf r})dV}.
\eeq

To launch a GRB, a baryon-clean region above the disk is probably needed. 
This does not mean that a wider clean region is always better, since a
thick disk can help collimate the outflowing jet --- but we want to determine
whether such a region exists or not. To estimate the baryon-poor
opening above our disks, we define the opening angle $\theta_{\rm clean}$. 
This angle specifies the widest cone oriented along the BH spin in 
which the condition $\rho \leq \rho_{\rm cut}$ is everywhere
satisfied: if $\theta_{\rm clean}(r,\phi)$ is the opening
angle within which we have $\rho \leq \rho_{\rm cut}$ at radius $r$ and 
azimuthal coordinate $\phi$, 
then $\theta_{\rm clean}=\min_{r,\phi}{[\theta_{\rm clean}(r,\phi)]}$.
In these simulations, the numerical method
requires atmospheric corrections to be applied starting
at  $\rho = 6\times10^8{\rm g}/{\rm cm}^3$, so that it is impossible to
reliably predict the behavior of matter below that threshold.  Therefore,
we set $\rho_{\rm cut} = 3\times10^9{\rm g}/{\rm cm}^3$.

For precessing binaries we also compute the tilt $\beta$ and 
twist $\gamma$ of the disk,
as defined by Fragile \& Anninos~\cite{2005ApJ...623..347F}. 
If $x^{\mu}$
are the inertial coordinates, $T^{\mu\nu}$ the stress-energy tensor, 
$\epsilon_{\mu \nu \sigma i}$ the Levi-Cevita tensor (with its last index
limited to nonzero values),
${\bm J}_{\rm BH}$ the angular momentum of the BH and $\hat{e}_y$ an arbitrary
unit vector orthogonal to ${\bm J}_{\rm BH}$, then $\beta$, $\gamma$ and
the disk angular momentum ${\bm J}_{\rm disk}$ are given by
\beqn
L^{\mu\nu} &=& \int \left(x^{\mu}T^{\nu 0} - x^{\nu}T^{\mu 0}\right)d^3x\\
S^{\mu} &=& \int T^{\mu 0} d^3x \\
J_{{\rm disk},i} &=& \frac{\epsilon_{\mu \nu \sigma i}L^{\mu\nu}S^{\sigma}}
{2\sqrt{-S^{\alpha}S_{\alpha}}}\\
\beta(r) &=& \arccos{\left[\frac{{\bm J}_{\rm BH}.{\bm J}_{\rm disk(r)}}
{|{\bm J}_{\rm BH}||{\bm J}_{\rm disk(r)}|}\right]} \\
\gamma(r)&=& \arccos{\left[\frac{{\bm J}_{\rm BH}\times{\bm J}_{\rm disk(r)}}
{|{\bm J}_{\rm BH}\times{\bm J}_{\rm disk(r)}|}.\hat{e}_y\right]}.
\eeqn
These parameters determine the 
inclination and the precession of the disk with respect to the spin of 
the black hole:
if ${\bf J}_{\rm BH}=J_{\rm BH}\hat{e}_z$, then
the orbital angular momentum of the disk at radius $r$ is written as 
\beq
{\bf J}_{\rm disk}
(r)=J_{\rm disk}(r)\left(\sin{\beta}\cos{\gamma} \hat{e}_x 
+ \sin{\beta}\sin{\gamma} \hat{e}_y + \cos{\beta}\hat{e}_z\right).
\label{eq:jdisk}
\eeq

Another useful property of the disk is its scale height, $H$. For a disk with
exponentially decreasing density, $H$ is defined by $\rho=\rho_c e^{-z/H}$.
Here, however,
the vertical profile of the disk is significantly more complex, and various
definitions of $H$ could be considered. We use the spread of the density
distribution $\rho(\theta,\phi)$ on a sphere of constant radius $r$ with its
polar axis along ${\bf J}_{\rm disk}(r)$ and define
\beq
H(r) = r \tan^{-1}\left({\sqrt{\mu \frac{\int \rho({\bf r}) 
[\theta({\bf r})]^2\, dS}{\int \rho({\bf r})\, dS}}}
\right).
\eeq
The parameter $\mu$ is somewhat arbitrary. For an exponential profile 
$\mu \sim 0.5$, while for a constant density profile ($\rho=\rho_0$ for 
$\theta < H/r$ and $\rho=0$ otherwise) we have $\mu \sim 3$. The disks observed
in our simulations are somewhat in between these two extremes. Accordingly,
we make the approximate choice $\mu=1$.  

To measure the accuracy of our simulations, we monitor the ADM Hamiltonian
and momentum constraints, and the generalized harmonic
constraints $\|\mathcal{C}\|$~\cite{Lindblom:2006}. At our
middle resolution, $\|\mathcal{C}\|$ peaks below $1\%$ for all cases and 
is less than $0.1\%$ during most of the inspiral. We also monitor
the ADM mass $M_{\rm ADM}$ and angular momentum ${\bf J}_{\rm ADM}$.  An
important check of our simulations is that the changes in these quantities
match the flux of energy and angular momentum in the outgoing gravitational
radiation, which we reconstruct from the Newman-Penrose scalar
$\psi_4$ as in Ref.~\cite{Boyle2007}.

\section{Cases}
\label{cases}

In order to assess the influence of the black hole spin on the disruption and
merger of BHNS binaries, we study configurations for which all
other physical parameters are held constant. The mass of the black hole is
$M_{\rm BH}=3M_{\rm NS}$, where $M_{\rm NS}$ is the ADM mass of an isolated
neutron star with the same baryon mass as the star under consideration, and
the initial coordinate separation is $d=7.5M$, with $M=M_{\rm BH}+M_{\rm NS}$.
For the nuclear equation of state, we use the polytrope
\beq
P = (\Gamma - 1)\rho\epsilon = \kappa \rho^\Gamma + {\overline T}\rho \,
\eeq
where ${\overline T}$ is a fluid variable related to, but not equal to,
the physical temperature.  We set $\Gamma=2$ and choose $\kappa$ so that the
compaction of
the star is $C=M_{\rm NS}/R_{\rm NS}=0.144$. For polytropic equations of state,
the total mass of the system does not have to be fixed: results can easily be
rescaled by $M$ (see e.g. Sec. II-F of 
Foucart {\it et al.}~\cite{Foucart:2008a}). 
However, whenever we choose to interpret our results in physical units 
(ms, km, $M_\odot$), we will assume that $M_{\rm NS}=1.4M_\odot$ 
($M=5.6M_\odot$). For that choice, the neutron star has a 
radius $R_{\rm NS}=14.6{\rm km}$, and the initial separation is $d=63{\rm km}$.

The different initial configurations and black hole spins studied are summarized
in Table \ref{table:init}. We consider 3 different magnitudes of the
dimensionless spin $a_{\rm BH}/M_{\rm BH}=(0,0.5,0.9)$,
all aligned with the orbital angular momentum. Then, we vary the inclination
angle $\phi_{\rm BH}$ between the spin of the black hole and the initial
angular velocity of the system, ${\bm \Omega}^{\rm rot}$. Considering that
most BHNS binary systems are expected to have $\phi_{\rm BH}\leq 90^\circ$
(Belczynski et al.~\cite{2008ApJ...682..474B}), with about half of the binaries
at $\phi_{\rm BH}\leq 40^\circ$, we choose 
$\phi_{\rm BH}=(20^\circ,40^\circ,60^\circ,80^\circ)$. The orientation
of the component of the BH spin lying in the orbital plane could 
also have measurable consequences. For example, 
Campanelli {\it et al.}~\cite{2007PhRvL..98w1102C} showed that the superkick 
configuration found in binary black hole systems is sensitive to the 
direction of the misaligned component of the BH spin. For BHNS
binaries, kicks are relatively small, and we are more interested in
the characteristics of the final black hole-disk system. After looking
at different orientations for $\phi_{\rm BH}=80^\circ$, we find that
the influence of the orientation of the misaligned component of the BH spin
is negligible compared to the influence of $\phi_{\rm BH}$. For this
first study of misaligned spins, we will thus limit ourselves to configurations
for which the initial spin lies in the plane generated by the initial orbital
angular momentum and the line connecting the two compact objects. 
As the different initial configurations do not use the same
background metric, there is no guarantee that two binaries with the same
initial coordinate separation can be directly compared. A better comparison
between initial configurations is the orbital angular velocity of the system.
In Table~\ref{table:init}, we show that all configurations have initial
angular velocity within $1\%$ of each other. This is the level of error
that we expect from the quasiequilibrium method for binaries at this
separation.

\begin{table}
\begin{tabular}{ccccccccc}
\hline \hline  
\tstruta Case  &
$a_{\rm BH}/M_{\rm BH}$  &
$\phi_{\rm BH}$  &
$\Omega_{\rm init}M$  &
$E_b/M_{\rm ADM}$  &
$J_{\rm ADM}/M_{\rm ADM}{}^2$ &
$t_{\rm merger}$
\\ \hline
\tstrutb s0      & 0   & -   & 4.16e-2  & 9.5e-3  & 0.66 & 7.5ms\\
s.5i0   & 0.5 & 0   & 4.11e-2  & 1.01e-2  & 0.91 & 11.4ms\\
s.9i0   & 0.9 & 0   & 4.13e-2  & 9.6e-3  & 1.13 & 15.0ms\\
s.5i20  & 0.5 & 20  & 4.09e-2  & 1.01e-2  & 0.90 & 10.5ms\\
s.5i40  & 0.5 & 40  & 4.10e-2  & 1.01e-2  & 0.87 & 9.9ms\\
s.5i60  & 0.5 & 60  & 4.11e-2  & 9.9e-3  & 0.82 & 9.0ms\\
s.5i80  & 0.5 & 80  & 4.13e-2  & 9.6e-3  & 0.76 & 7.7ms\\
\hline \hline  
\end{tabular}
\caption{
  Description of the cases evolved.  $a_{\rm BH}/M_{\rm BH}$ is the
initial dimensionless spin of the BH, $\phi_{\rm BH}$
is its inclination with respect to the initial orbital angular
momentum and  $E_b$ is the initial binding energy. $t_{\rm merger}$
is defined as the time by which half of the matter has been accreted
by the BH. Differences in the initial angular velocity
and binding energy are within the margin of error of the initial data:
at this separation the eccentricity reduction method can require variations 
of $\Omega_{\rm init}$ of order $1\%$, and modifies the binding energy by 
a few percent.
}
\label{table:init}
\end{table}

\section{The nonspinning case:  a test of our accuracy}
\label{nonspinning}

As an example, we consider the case s0, in which the BH is initially
nonspinning.  We evolve this case at each of our three resolutions. 
After a short (two orbits) inspiral, the neutron star is disrupted, and
most of the matter is quickly swallowed by the black hole.  The remainder
expands into a tidal tail and then falls back to form an accretion disk. 

Nearly identical systems have been studied both by
Shibata~{\it et al}~\cite{Shibata:2009cn} and by
Etienne~{\it et al}~\cite{Etienne:2008re}.  The former found an
insignificant disk after merger, while the latter found 4\% of the NS
mass still outside the hole 300$M$ ($\sim 8{\rm ms}$) after merger.  
Both groups found
a final BH spin of $s=0.56$.  We find a disk mass of 3.7\% at 300$M$ after
merger, smaller than in~\cite{Etienne:2008re}, but closer to this result
than to that in~\cite{Shibata:2009cn}.  Our final BH spin is 0.56, in
agreement with both previous studies.

\begin{figure}
\includegraphics[width=8cm]{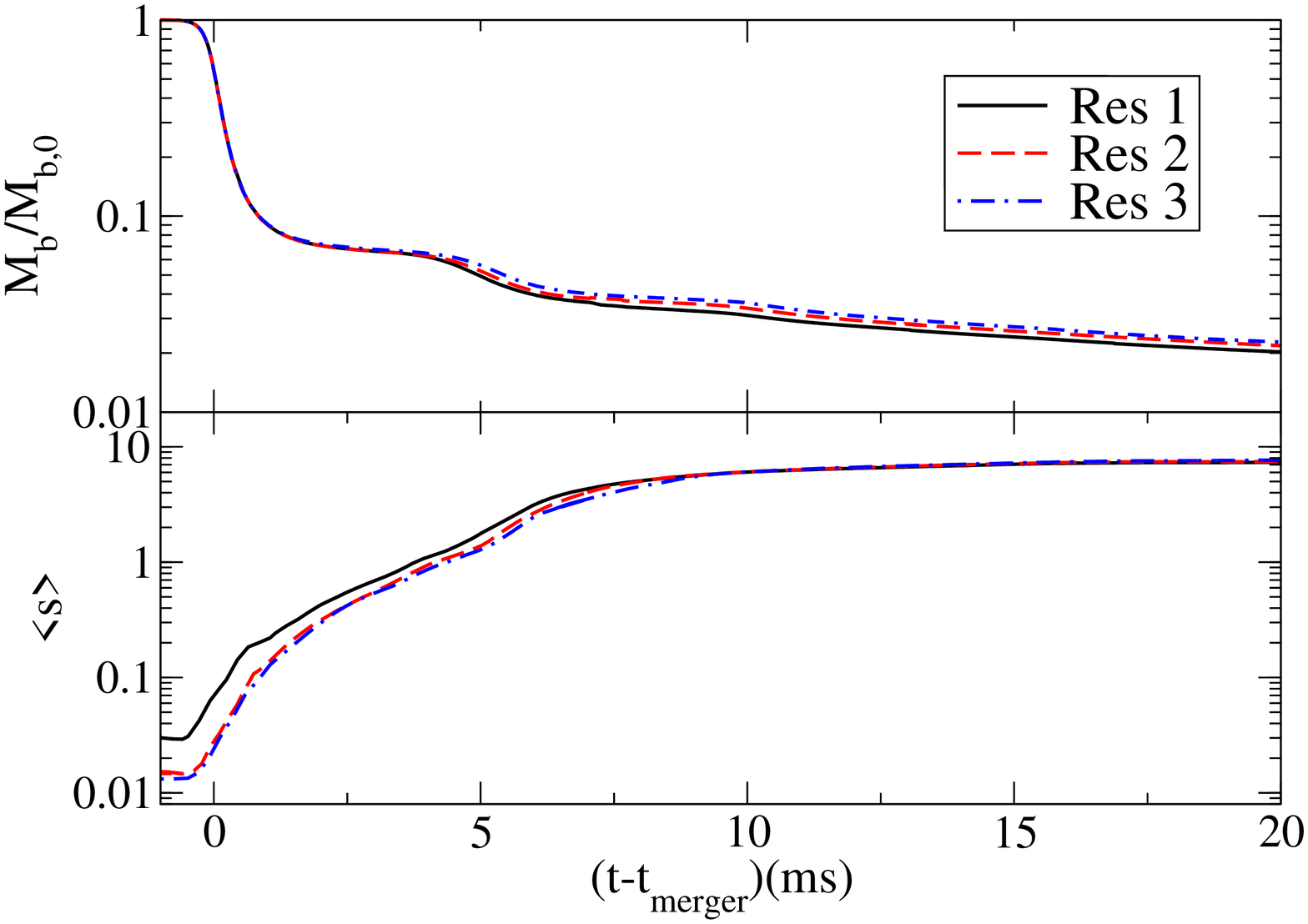}
\caption{
  Baryonic mass $M_b$ normalized by its initial value $M_{b,0}$ 
and average entropy $\langle s \rangle$ for three resolutions.
}
\label{fig:matter_convg}
\end{figure}

In Fig.~\ref{fig:matter_convg}, we show the evolution of $M_b$ and 
$\langle s\rangle$ for the entire merger phase for the three resolutions. 
Reassuringly, the different resolutions give very similar results, with
the two higher resolutions being closest together.  The baryonic mass is
initially constant before accretion starts.  Then, as the NS is disrupted
and the core of the star is swallowed, $M_b$ drops rapidly.  It next levels
off while the remaining matter is in an accretion disk and an expanding tidal
tail.  When the tidal tail falls back onto the disk, there is a
second phase of rapid accretion, after which the accretion rate settles down
to a low
value.  At the end of the simulation, the accretion time scale is
$\tau_{\rm disk}\sim 55{\rm ms}$, implying that
the total lifetime of the disk would
be around 75ms.  
At late times, the deviation in $M_b$ between resolutions
becomes somewhat larger, indicating that our errors have
accumulated to about $0.1\%$ of the initial mass.  
For the purposes
of this paper, this is adequate, since the disk on these time scales is
affected by magnetic and radiation processes not included in the simulations. 
However, future long-term disk simulations will require higher accuracy.

As for the entropy, at the beginning of the merger it only deviates from zero 
because of numerical heating during the inspiral.  As expected, this numerical
heating is significantly lower at higher resolutions.  The post-merger
heating is not numerical, but a physical consequence of shocks in the
disk and the disk-tail interface.  A confirmation that the heating is
physical is that it is nearly the same for all resolutions, and its
magnitude is much larger than the numerical heating.  When the disk settles,
there is no further shock heating, so the entropy levels off.  This indicates
that the heating due to numerical viscosity is small compared to shock
heating.  Unfortunately, this is not the same as saying that the
numerical viscosity is irrelevant altogether.  However, the closeness of
$\tau_{\rm disk}$ at each resolution indicates that this viscosity is not
the main driving force of the accretion.  The average temperature
$\langle T\rangle$ behaves in a way similar to the entropy.
Starting from low
values, it increases after the merger and stabilizes around 3MeV. 
All resolutions show the same $\langle T\rangle$ growth, and all level off
at the same value.  After leveling off, though, $\langle T\rangle$ displays
0.1MeV oscillations that do not converge well, another indication that
our accuracy is sufficient for some but not all purposes.

\begin{figure}
\includegraphics[width=8cm]{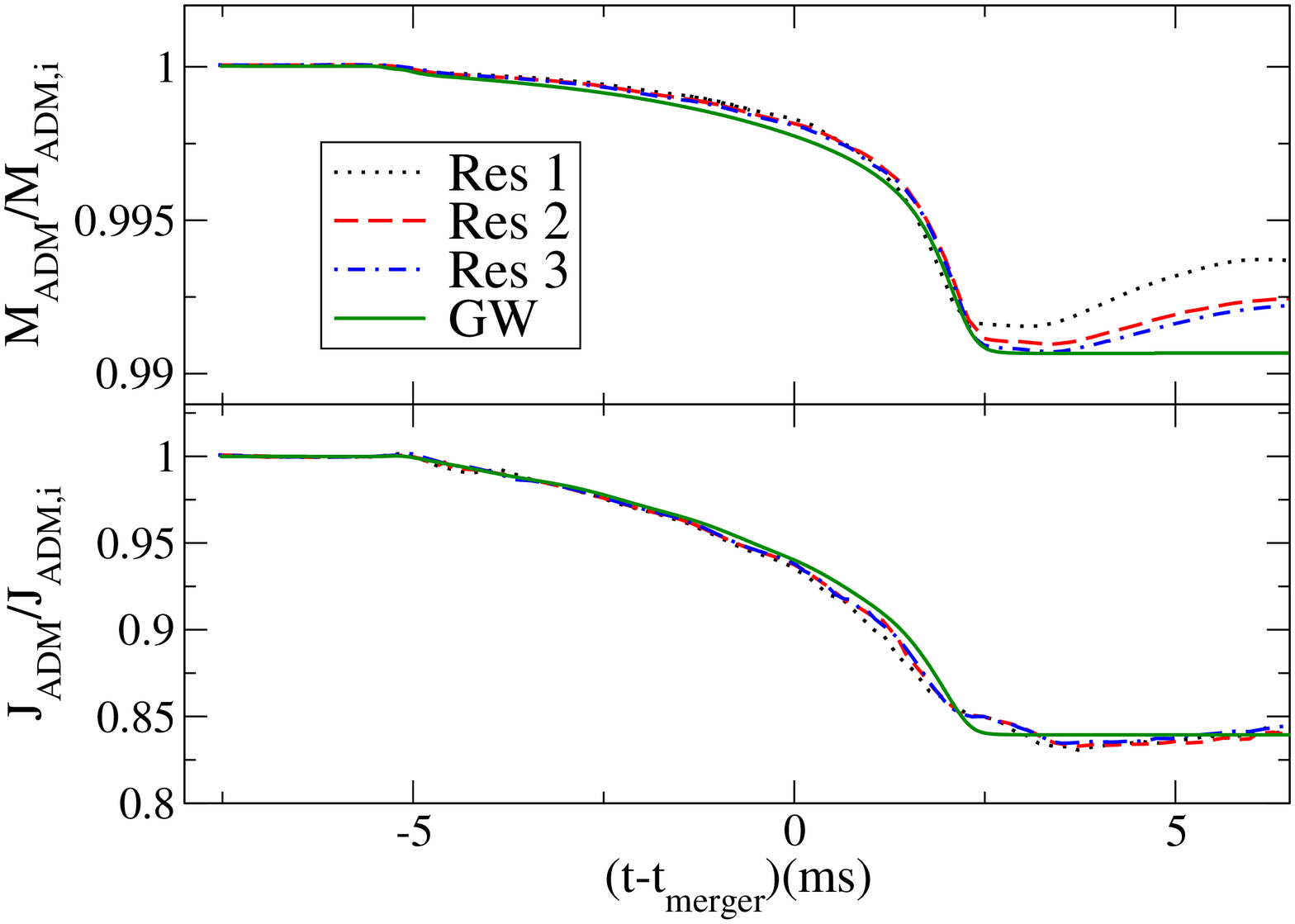}
\caption{
  $M_{\rm ADM}$ and $J_{\rm ADM}$ (normalized to their initial
values) compared to the changes expected
  from the gravitational radiation flux for three resolutions.
}
\label{fig:ADM_MJ_convg}
\end{figure}

In Fig.~\ref{fig:ADM_MJ_convg}, we plot the ADM energy and orbital-axis
angular momentum measured on a surface $75M$ from the center of mass of the
system.  Also plotted
is the evolution of these quantities expected from the gravitational radiation
through this surface.  Overall, the agreement is quite good, although there
is some deviation in $M_{\rm ADM}$ a while after the merger.  This seems
to be associated with an increase in constraint violations at the merger
time.  The relative constraint violations, as measured by $\|\mathcal{C}\|$,
peak slightly below 1\% at the middle resolution. The corresponding values
for the ADM constraints are 1 --2 \%,
before both constraints fall back to low values. The deviations in 
$M_{\rm ADM}$ happen around the time this constraint-violating pulse reaches 
the $r=75M$ surface.

We have checked several other quantities, including the black hole mass and
spin and the gravitational waveform.  All of these show very good convergence.

\section{Results}
\label{results}

The general behavior of our simulations is typical of BHNS binaries
for which the NS is disrupted outside
the innermost stable circular orbit
of the BH. From an initial separation of $60{\rm km}$, the compact objects
go through 2-3 orbits of inspiral driven by the emission of gravitational
waves. When the distance between them has been reduced to about 
$30-40{\rm km}$,
tidal forces cause the neutron star to disrupt. Most of the matter is rapidly
accreted into the black hole, while the rest is divided between a long tidal
tail, expanding about $200 {\rm km}$ away from the black hole, 
and a developing accretion disk. The duration
of the inspiral varies with the spin of the BH, with the component of the BH
spin along the orbital angular momentum delaying the merger. The merger
time $t_{\rm merger}$, which we define as the time at which 50\% of the matter
has been accreted onto the BH, is listed for all cases in 
Table~\ref{table:init}.

The resulting accretion disk is highly asymmetric, and evolves in time. 
When the disk first forms, around 5ms after merger, it
creates a torus of matter with its peak surface density 
(baryon density integrated over the height of the disk) at 
$r(\Sigma_{\rm max})\sim 30{\rm km}$ 
and a temperature $T\sim 1-2{\rm MeV}$. 
Then, as matter accretes from the tidal tail and shocks heat the fluid,
the disk expands quickly. About $10-20{\rm ms}$ after merger, the disk starts
to settle into a stable, slowly accreting state. To compare the
different configurations studied here, we look at the properties of this late-time 
stable configuration, listed in Table~\ref{table:disk}. In
Table~\ref{table:finalBH}, we give the characteristics of the final black
hole, as well as the kick velocity, the energy content of the emitted
gravitational waves, and the peak amplitude of a dominant (2,2) mode of 
the waves.  [The (2,2) and (2,-2) are the strongest modes, with nearly
equal amplitude.  See Section~\ref{spin_dir} on the higher modes.]

\begin{table}
\begin{tabular}{ccccccccc}
\hline \hline  
\tstrut Case  &
$M_{\rm disk}/M_{\rm NS}$ &
$\langle T\rangle_{\rm disk}$ &
$\beta(\Sigma_{\rm max})$ &
$r(\Sigma_{\rm max})$ &
$\theta_{\rm clean}$ &
$\frac{H}{r}(\Sigma_{\rm max})$ &
\\ \hline  
\tstrutb s0      & 5.2\%  & 3.0MeV & $0^\circ$  & 50km  & $50^\circ$ & 0.20 \\
s.5i0   & 15.5\% & 3.5MeV & $0^\circ$  & 50km  & $35^\circ$ & 0.25 \\
s.9i0   & 38.9\% & 5.6 MeV & $0^\circ$  & 20km  & $8^\circ$ & 0.18 \\
s.5i20  & 14.5\% & 3.6MeV & $2^\circ$  & 50km  & $40^\circ$ & 0.20 \\
s.5i40  & 11.5\%  & 3.8 MeV & $4^\circ$  & 50km & $30^\circ$ & 0.22 \\
s.5i60  & 8.0\%  & 3.7MeV & $7^\circ$ & 50km  & $40^\circ$ & 0.20 \\
s.5i80  & 6.1\%  & 3.6MeV & $8^\circ$ & 50km  & $50^\circ$ & 0.25 \\
\hline \hline  
\end{tabular}
\caption{
  Properties of the accretion tori at late time. The mass of the disk 
$M_{\rm disk}$ (baryon mass outside the excision region), 
which decreases continuously due to accretion onto the BH, is measured at 
$t_{\rm merger}+5{\rm ms}$. Even at late times, all quantities still show
oscillations of $\sim 10\%$.\\
} 
\label{table:disk}
\end{table}

\begin{table}
\begin{tabular}{cccccccc}
\hline \hline  
\tstruta Case  &
$M_{\rm BH}/M$ &
$a_{\rm BH}/M_{\rm BH}$   &
$v_{\rm kick}{\rm(km/s)}$  &
$E_{\rm GW}/M$   &
$rM\Psi_4^{2,2}$ &
\\ \hline  
\tstrutb s0      & 0.97  & 0.56  & 53 & 0.98\%  & 0.020  \\
s.5i0   & 0.94  & 0.77  & 60 & 0.92\%  & 0.012  \\
s.9i0   & 0.89  & 0.93  & 52 & 0.95\%  & 0.009  \\
s.5i20  & 0.95  & 0.76  & 60  & 0.89\%  & 0.012 \\
s.5i40  & 0.96  & 0.74  & 61  & 0.91\%  & 0.013 \\
s.5i60  & 0.96  & 0.71  & 54  & 0.95\%  & 0.014  \\
s.5i80  & 0.97  & 0.66  & 67  & 0.95\%  & 0.017  \\
\hline \hline  
\end{tabular}
\caption{
  Properties of the post-merger black hole and gravitational waves.
\label{table:finalBH}
}
\end{table}

\subsection{Effects of spin magnitude}
\label{spin_mag}

To test the effects of BH spin magnitude, we compare our results for
s0, s.5i0, and s.9i0.  A comparison of this type has already been performed
by Etienne~{\it et al}~\cite{Etienne:2008re}.  Our cases are different, though:
unlike them, we do not consider an antialigned case, but we do push the BH
spin to a slightly higher level in our run s.9i0.

Run s.9i0 presented special numerical challenges. In {\sc spec},
the singularity and inner horizon of the BH have to be excised from the
numerical grid while the apparent horizon must remain outside the excision
surface. But for nearly extremal black holes the region between the inner 
horizon and the apparent horizon becomes very narrow. To perform excision in 
such cases, the excision  boundary must nearly
conform to the apparent horizon.  We do this by introducing a coordinate
map in the initial data so that the horizon is initially spherical on the
pseudospectral grid.  We then use our dual frame coordinate-control
method~\cite{2006PhRvD..74j4006S}
to fix the location of the horizon throughout the whole simulation.  For
lower spins this is not necessary, and we only begin to control the horizon 
location at the time of neutron star disruption. That modification excepted, 
case s.9i0 was simulated in exactly the same way as the other cases.  
The deviation between the results at resolutions
Res2 and Res3 is somewhat larger than in the other cases 
(though still quite small).  To be more exact, the relative deviation 
in the disk mass between
Lev2 and Lev3 is about 9\% ($\sim$3\% of the NS mass) 
for s.9i0 while it was about 5\% for s0,
and the deviation in merger time is about 4\% for s.9i0 but 
only 0.3\% for s0.  The difference indicates that high resolution is needed 
when studying such extreme cases.

We find that systems with higher $a_{\rm BH}/M_{\rm BH}$ spiral in more slowly: 
from the same initial separation, s0, s.5i0, and s.9i0 take roughly 2, 3,
and 3.7 orbits, respectively, before NS disruption begins.  This effect
exists in the post-Newtonian 
treatment~\cite{1995PhRvD..52..821K} and it has already been seen both in 
binary black hole
(e.g.~\cite{2006PhRvD..74d1501C}) and BHNS~\cite{Etienne:2008re}
simulations.  Because of the prolonged inspiral for the high-spin cases,
more angular momentum is radiated:  0.11$M^2$ for s0 vs 0.14$M^2$ for s.9i0.
Additionally, as the BH becomes nearly extremal,
increasing the spin 
becomes more and more difficult. This is reflected in the final spin of the 
BH:  while $a_{\rm BH}/M_{\rm BH}$ increases from 0 to 0.56 for s0, it only 
increases from 0.9 to 0.93 for s.9i0. 

\begin{figure}
\includegraphics[width=8cm]{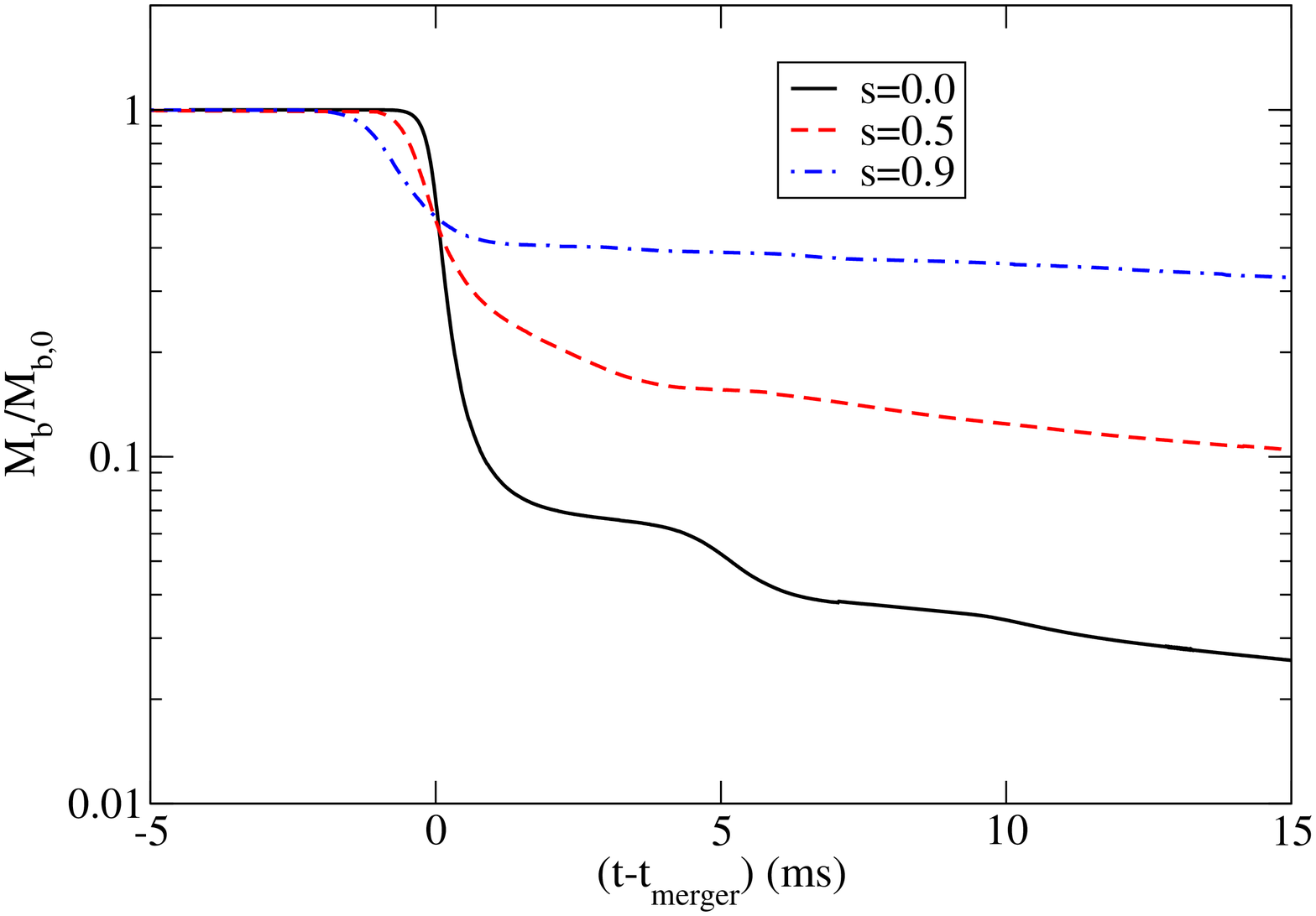}
\caption{
  Evolution of the total baryonic density outside the
  hole for three different aligned BH spins.
}
\label{fig:mass_vs_spin_mag}
\end{figure}

We also confirm that the post-merger accretion disk mass
increases significantly as the magnitude of the aligned BH spin is increased,
as shown in Fig.~\ref{fig:mass_vs_spin_mag}. 
This is in qualitative agreement with Etienne~{\it et al}.
About $400M$(11ms) after merger they find $M_b/M_{b,i}\sim$ 20\% for an
$a_{\rm BH}/M_{\rm BH}=0.75$ system, while for our 
$a_{\rm BH}/M_{\rm BH}=0.9$ system, we find
$M_b/M_{b,i}\sim$35\% at a similar time.

\subsection{Effects of spin orientation}
\label{spin_dir}

Most BHNS binaries are expected to have at least a moderate
misalignment between the spin of the BH and the total
angular momentum~\cite{2008ApJ...682..474B}, and this should
affect all stages of the binary evolution. 

\begin{figure}
\includegraphics[width=7cm]{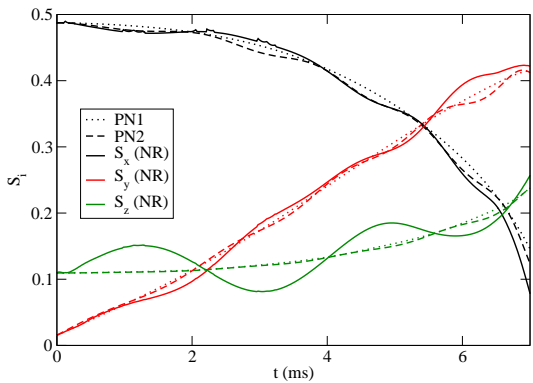}
\caption{
  Comparison of the evolution of the Cartesian
  components of the spin for an initial inclination
  of $80^{\circ}$ in our simulation (NR) and for
  the first- and second-order Post-Newtonian
  expansions (resp. 1PN and 2PN). The PN values
  are obtained by integrating the evolution
  equations for the spin given in \cite{2006PhRvD..74j4033F},
  using the trajectory and current spin of the
  numerical simulation.
}
\label{fig:prec}
\end{figure}

During inspiral, the orbital angular momentum and the
BH spin precess around the total angular momentum of the system.
The evolution of the coordinate components of the BH spin for the 
s.5i80 case is shown in Fig.~\ref{fig:prec}.
Over the two orbits of inspiral, the spin goes through about
a quarter of a precession period. The qualitative evolution of the spin
is well described by post-Newtonian corrections 
(see e.g.~\cite{2006PhRvD..74j4033F}),
even though our simulation uses a different gauge choice.
As for aligned spin, the infall velocity varies between cases:
the larger the component of the spin aligned with the angular
momentum, the slower the inspiral. Not too surprisingly, we find a monotonic
decrease of the merger time with increasing misalignment 
angle $\phi_{\rm BH}$, with $t_{\rm merger}(\phi_{\rm BH},a_{\rm BH}) 
\rightarrow t_{\rm merger} (0,0)$ for $\phi_{\rm BH} \rightarrow 90^\circ$.

\begin{figure*}
\begin{tabular}{cc}
\includegraphics[width=9cm]{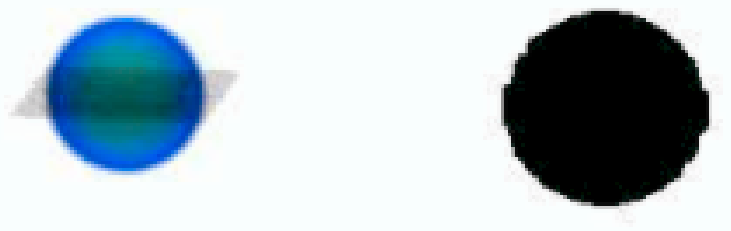} &
\includegraphics[width=9cm]{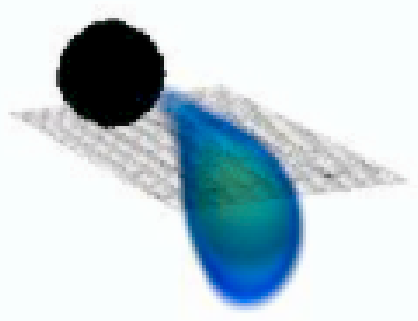} \\
\includegraphics[width=9cm]{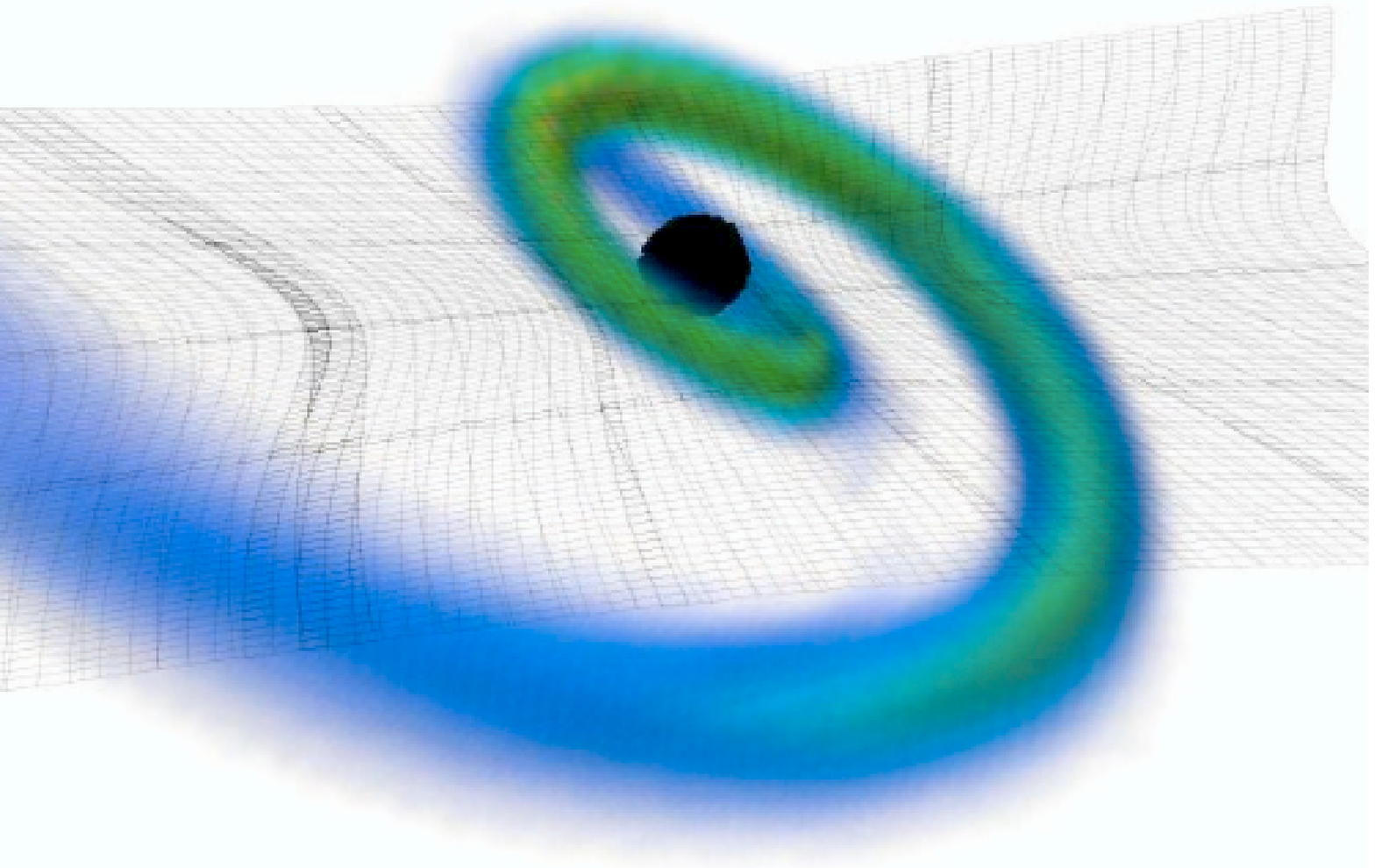} &
\includegraphics[width=9cm]{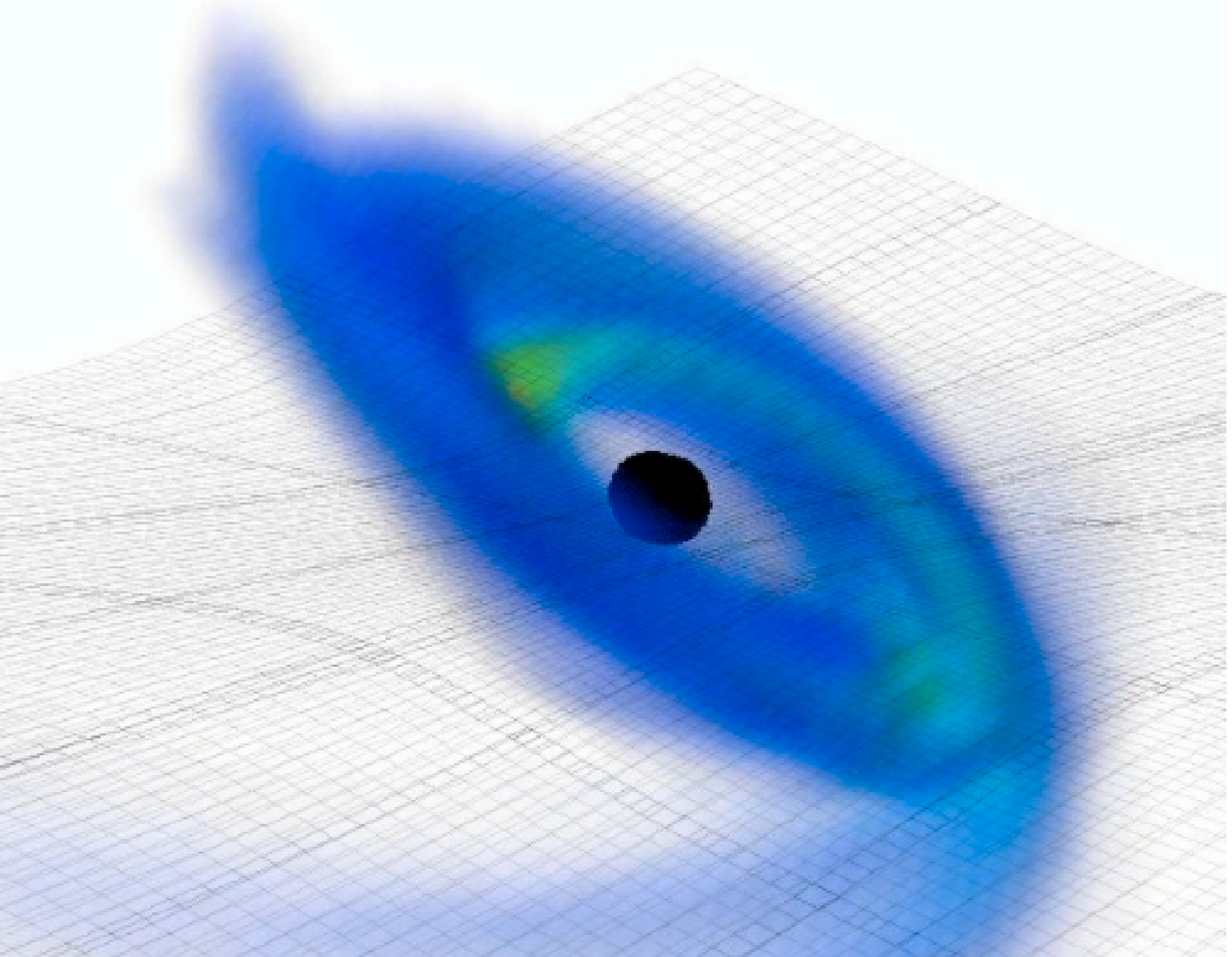} \\
\end{tabular}
\caption{Evolution of an inclined binary (s.5i80). {\it Top left:} Beginning of
the simulation, at a separation of 63km. {\it Top right:} After 7ms and two
orbits of inspiral, the star disrupts and most of the matter rapidly accretes 
onto the black hole. {\it Bottom left:} After 11ms, the remaining matter 
($\sim 8.5\%$ of the initial mass) is split between a developing 
disk and a tidal tail.
Differential precession between the disk and the tail means that the disk and
the matter falling back from the tail orbit in different planes. {\it Bottom
right:} After 15.5ms, the disk contains about $4.5\%$ of the initial mass of
the star. It is still highly inhomogeneous, and slowly expanding. A movie
of the whole simulation is available online~\cite{Movie}. In each image, the
wired frame shows the ``fluid'' grid in the $z=0$ plane 
(orbital plane at $t=0$).}
\label{fig:movie}
\end{figure*}

The disruption of the star and formation of a disk, shown in 
Fig.~\ref{fig:movie}, proceed 
somewhat differently from what is observed for nonprecessing
binaries. As before, the disruption of the star is accompanied by
the formation of a long tidal tail. But because of the inclination 
of the BH spin, the orbital plane of the fluid continues to 
precess after disruption. Because the precession rate varies with the
distance to the hole, the tail and the disk do not remain
in the same plane. While for nonprecessing binaries matter from
the tail falls back within the orbital plane of the disk, 
here matter is added to the disk at an angle varying in
time. This significantly modifies the nature of the tail-disk
interactions. At small inclination angles 
($\phi \sim 20-40^\circ$), we have a direct collision 
between the developing disk and the tidal tail, while for larger inclinations
the disk is initially formed of layers of high-density material at different 
angles with respect to the black hole spin. 

\begin{figure}
\includegraphics[width=7cm]{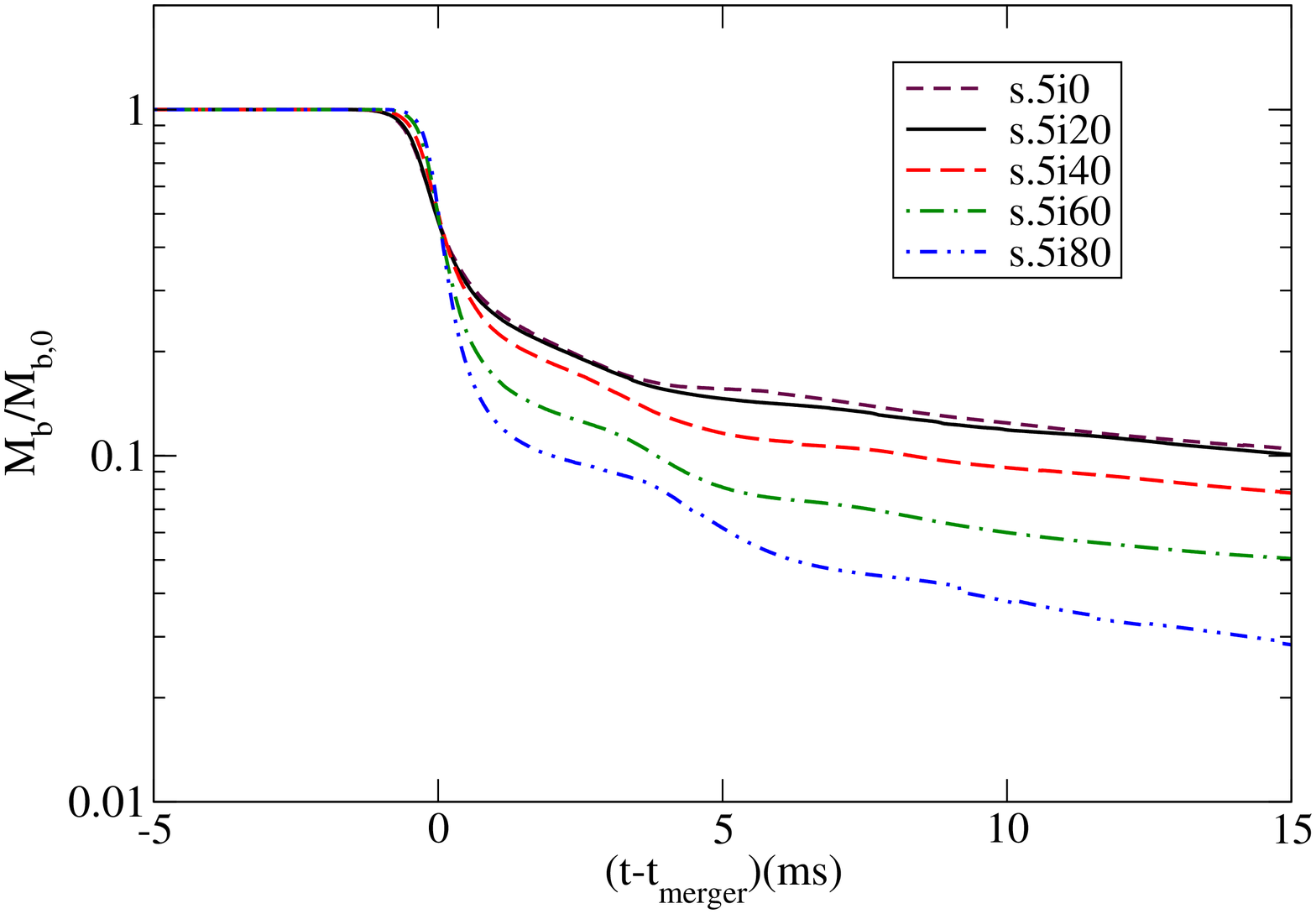}
\caption{
  Evolution of the total baryonic mass outside the
  hole for different initial inclination of the BH
  spin.
}
\label{fig:masscomp}
\end{figure}

The mass of the disk, plotted in Fig.~\ref{fig:masscomp}, decreases as 
the inclination of the binary increases. The transition between low and 
high mass disks is continuous, but more rapid at large inclinations: for
$\phi_{\rm BH}<40^\circ$, $10-15\%$ of the initial mass of the
star ($\sim 0.15-0.2M_\odot$) remains 
either in the tail or in the disk 5ms after merger. This is
roughly similar to the disk formed for $\phi_{\rm BH}=0$. At
higher inclinations, the size of the disk drops sharply, to
about $5\%$ of the initial mass of the star. We expect the
disk mass to be even lower for antialigned spins 
($\phi_{\rm BH}>90^\circ$), though such configurations appear
less likely to be found in astrophysical systems. These changes 
in the disk mass with the orientation of the BH spin show some
similarities with the results of 
Rantsiou {\it et al.}~\cite{2008ApJ...680.1326R}, obtained in the 
small mass-ratio limit ($q = 1/10$) by using a static
background metric. They found that for a disk to be formed,
the condition $\phi_{\rm BH} < 60^\circ$ has to be satisfied. As our
simulations use a mass ratio $q=1/3$, which is more
favorable to the formation of a disk, it is not too surprising
that even high inclinations leave us with a significant disk;
however, the influence of inclination remains important for 
$\phi_{\rm BH} > 40^\circ$. These factors are particularly useful 
when considering the potential of the final remnant to be a
short gamma-ray burst
progenitor. Since the influence of a misaligned BH spin is only felt for
$\phi_{\rm BH} > 40^\circ$, the majority of BHNS binaries can form disks
about as massive as is predicted by simulations that do not take into account
the inclination of the BH spin. 

\begin{figure}
\includegraphics[width=7cm]{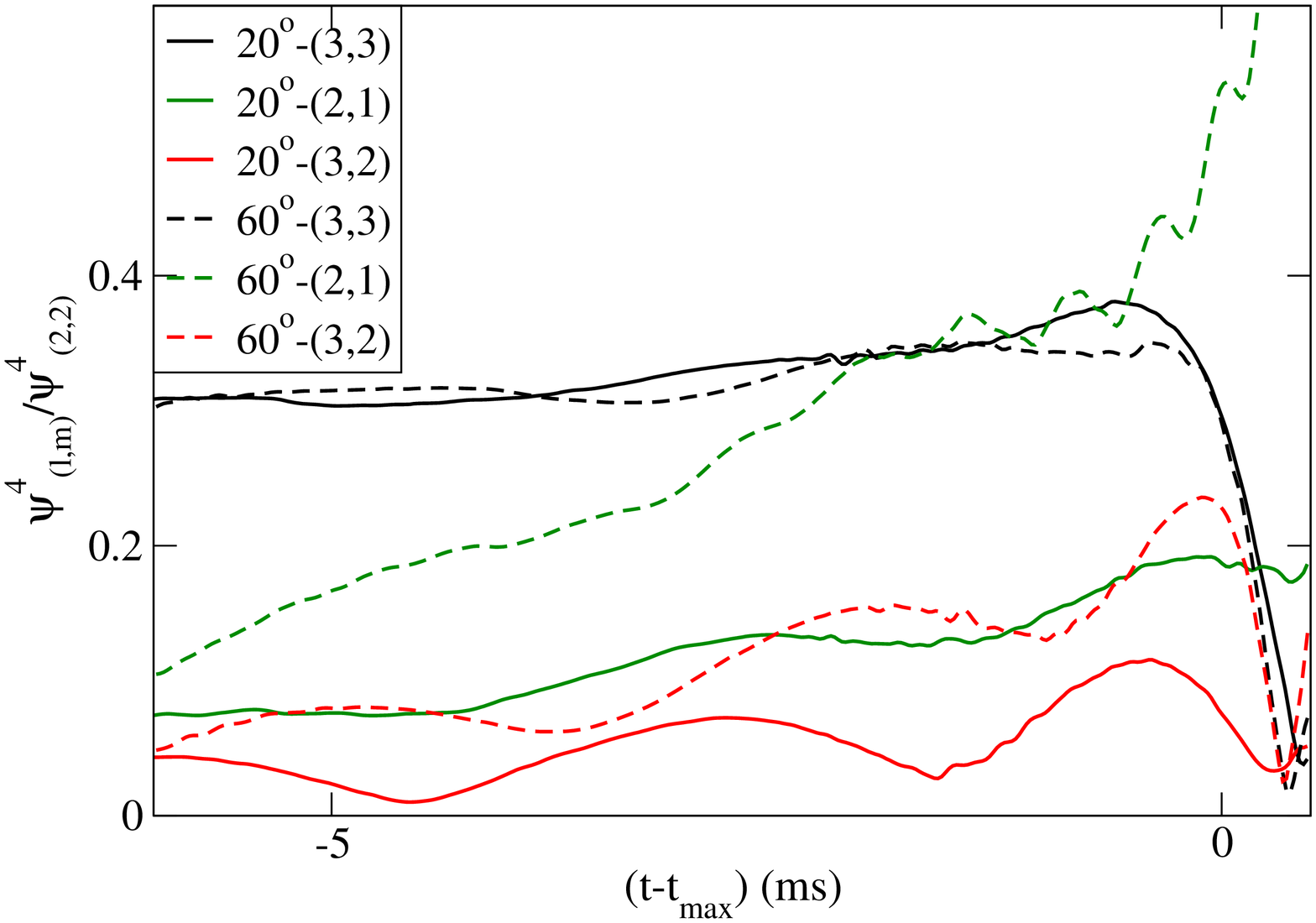}
\caption{
  Amplitude of $\psi^4_{(l,m)}/\psi^4_{(2,2)}$ for the 3 modes (2,1), (3,3) and
  (3,2), shown for initial inclinations of the BH spin of $20^\circ$ and 
  $60^\circ$. The time $t_{\rm max}$ corresponds to the peak 
amplitude of the dominant (2,2) mode.
}
\label{fig:HighModes}
\end{figure}

The gravitational wave signal is also significantly affected by the
value of $\phi_{\rm BH}$. We expand the Newman-Penrose scalar $\Psi_4$ 
extracted at 
$R=75M$ using the spin-weighted spherical harmonics $_{-2}Y_{lm}$
and choosing the polar axis along the initial orbital angular momentum of the 
binary. 
The peak amplitude of the dominant (2,2) mode of $\Psi_4$
will increase for large 
inclinations, as could be expected from the results
obtained for aligned spins; here too, a large component of the spin along 
the orbital angular momentum works against large $\Psi_4$
amplitudes.
Additionally, the contribution of subdominant modes can become significant 
at large inclinations. In Fig.~\ref{fig:HighModes}, we show the ratio of the 
amplitude of the scalar $\Psi_{4(l,m)}$ to the amplitude
of the (2,2) mode for various $(l,m)$ modes and
for $\phi_{\rm BH}=20^\circ,60^\circ$. The modes
most strongly affected by the precession of the binary are the (2,1) and (3,2)
modes, the first reaching half the amplitude of the dominant mode around
merger for the s.5i60 simulation. Analytical predictions for the
effect of a precessing trajectory on the modal decomposition of the 
gravitational wave signal have been derived by 
Arun {\it et al.}~\cite{2009PhRvD..79j4023A}. We find qualitative 
agreement with their results if we assume that the 
compact objects follow the trajectories 
obtained from our numerical simulations. In particular, we note that for 
precessing binaries, the frequency of the (2,1) mode is closer to 
$2\Omega_{\rm rot}$ than to $\Omega_{\rm rot}$. The (2,1) and (2,2) modes
have similar frequencies, so that the ratio of their amplitude computed using
the scalar $\Psi_4$ is close to the result one would obtain by using the
gravitational strain $h$ instead. This will not be true for the (3,3) mode, which
has a frequency $\Omega_{(3,3)} \sim 3\Omega_{\rm rot}$:
since $\Psi_4 =\partial^2h/\partial t^2$, we have $h_{(3,3)}/h_{(2,2)} \sim (4/9)
\Psi_{4(3,3)}/\Psi_{4(2,2)}$. 

\begin{figure}
\includegraphics[width=7cm]{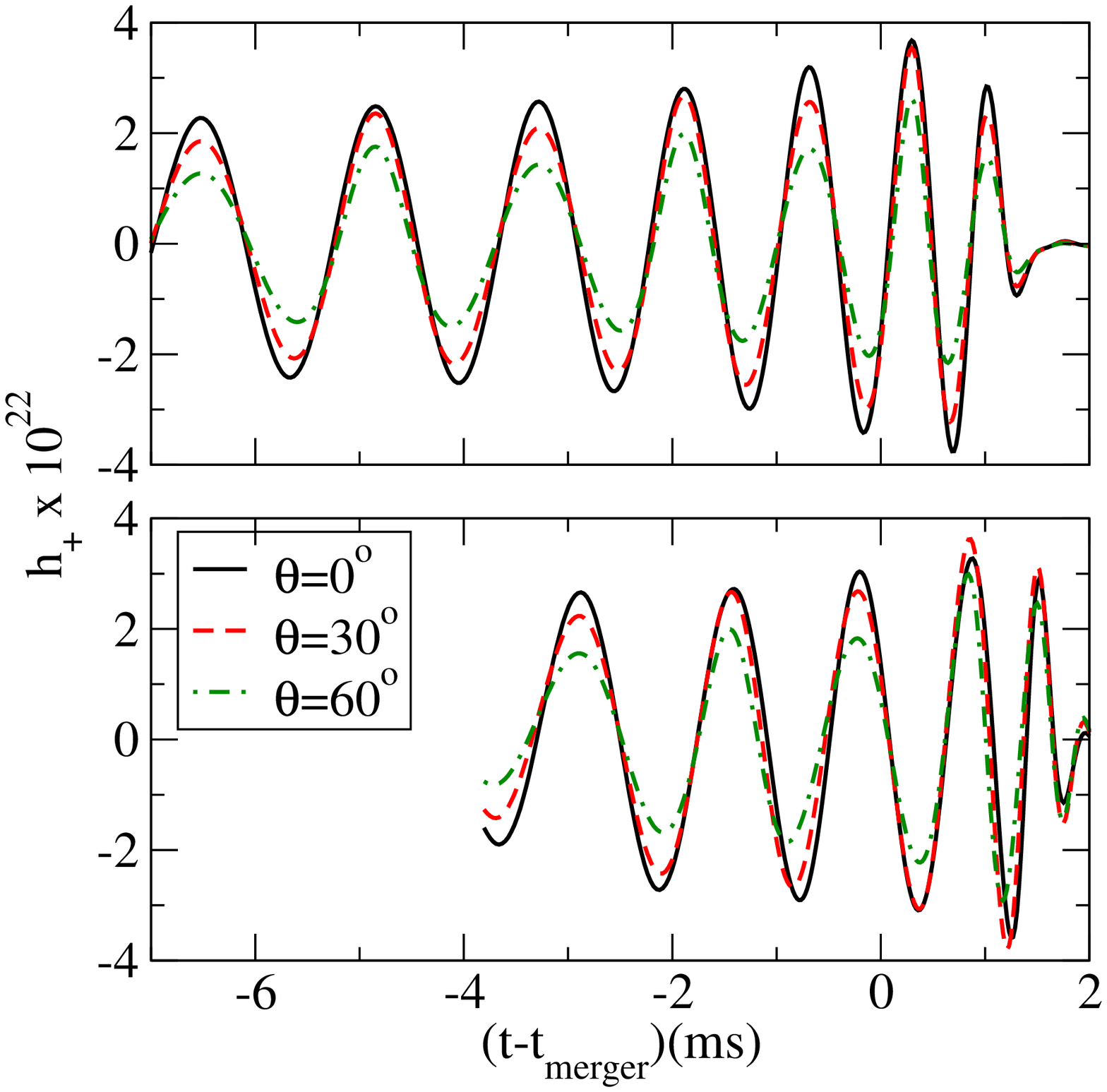}
\caption{
  Real part of the gravitational strain $h$ for the simulations
  s.5i0 ({\it top panel}) and s.5i80 ({\it lower panel}), 
viewed from different inclinations $\theta$ with respect
to the initial orbital angular momentum. The wave is extracted at $r=75M$ 
for a mass of the neutron star $M_{\rm NS}=1.4M_{\odot}$, and
assumed to travel as a linear perturbation up to the observation point located
at $r=100\,{\rm Mpc}$.
Waves emitted at the time of merger will reach the radius 
$r=75M$ at $(t-t_{\rm merger}) \sim 2{\rm ms}$.
Over the 2 -- 3 orbits simulated here,
the effects of the orbital precession---and, in particular, the contribution
on the second highest mode (2,1), shown in 
Fig.~\ref{fig:HighModes}---remain small.
}
\label{fig:hWithOrientation}
\end{figure}

In Fig.~\ref{fig:hWithOrientation}, we plot the gravitational
strain $h$ as observed from a distance of $100\,{\rm Mpc}$ for the simulations
s.5i0 and s.5i80. The three waveforms correspond to observation points
whose lines of sight are inclined by $0^\circ$, $30^\circ$ and $60^\circ$ with
respect to the initial orbital angular momentum of the system. Over the short
inspiral considered here, the effects of precession are relatively small. 
The main difference visible in these waveforms is the slower inspiral
experienced by the binary with aligned spin. We can also note that in the 
misaligned configuration the star does not disrupt as strongly as in the 
aligned case, causing the cutoff of the wave emission to occur at a later time.
As a consequence, the cutoff frequency of the wave is larger for misaligned
spins than for aligned spins. This explains why the amplitude of the 
gravitational strain $h$ is comparable for both configurations, while the 
misaligned case showed a significantly larger amplitude when the wave was
measured using the scalar $\Psi_4$ (a similar effect occurs if both spins are 
aligned but of different magnitudes). Finally, the precession of the orbit 
causes
the gravitational wave emission of the misaligned configuration to peak at 
a non-zero inclination with respect to the initial orbital angular momentum. 
Here, at the time of merger the wave measured at an inclination 
of $30^\circ$ is slightly larger than at $0^\circ$. For these
effects to be more visible, and in particular for a full precession period
to be observable, longer simulations are required ($\sim 10$ orbits).

\subsection{Post-merger accretion disks}
\label{disks}

The evolution of the accretion disk over time scales comparable to its
expected lifetime is likely to be significantly influenced by physical effects
that are not taken into account in our simulations, mostly the magnetic
effects and the impact of neutrino cooling. We do not expect the results
of our simulations to accurately represent the details of the late-time 
evolution of the disk, but we can nonetheless extract some information
regarding the general characteristics of the final remnant. To obtain these
approximate results, which are summarized in Table~\ref{table:disk}, we use 
the fixed-metric approximation described in Sec.~\ref{Cowling}, starting 
$5-10{\rm ms}$ after merger. At that time, the disk is still expanding,
and will typically settle down to a more stable quasiequilibrium profile
with a relatively low accretion rate over about $10{\rm ms}$. 

The coordinate distance between the peak surface density of the disk
(averaged over all points at a given coordinate radius) and the center of
the BH shows no strong or monotonic dependence on the BH spin. After the
initial expansion of the disk, variations in the details of the interactions 
between the tidal tail and the accretion disk can lead to different
evolutions of the density profile. On average, the disks tend to expand 
slightly, while their density decreases because of continued accretion onto the 
black hole. However, neither the tidal tail nor the disk are homogeneous, 
so that the evolution of the density profile shows significant oscillations
around that average behavior. The accretion rate is larger for the more massive
disks, so that the expected lifetime of the disk is of the same
order of magnitude for a nonspinning BH ($\tau \sim 75{\rm ms}$) as for
the highly spinning BH ($\tau \sim 150{\rm ms}$).

\begin{figure}
\includegraphics[width=7.3cm]{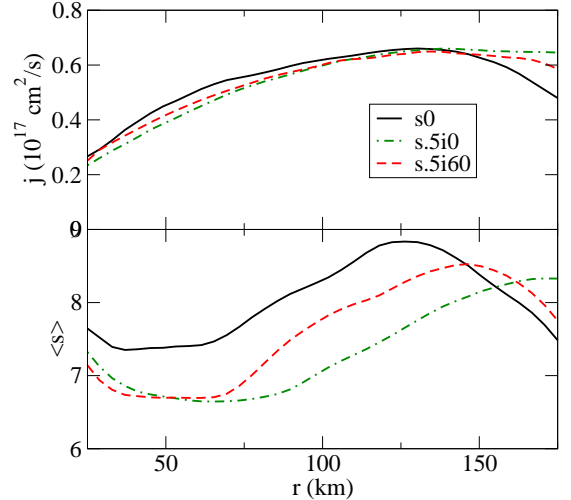}
\caption{
  {\it Upper panel:} Specific angular momentum of the accretion disk at 
$t=30{\rm ms}$ for runs s0, s.5i0 and s.5i60. Note that for the smaller
spin s0, the disk does not extend farther than $r \approx 125{\rm km}$.\\
{\it Lower panel:} For the same configurations, entropy of the disk averaged
over all points at a given distance from the black hole center.
}
\label{fig:profjands}
\end{figure}

The thermal evolution of the disk does not vary much between configurations.  
The temperature $\langle T\rangle_{\rm disk}$ rises rapidly during the 
formation of the disk, then stabilizes at about 3.5MeV in each case --- 
although the average entropy $\langle s\rangle_{\rm disk}$ is about 10\% lower 
for spinning black holes than for s0. For most configurations, the temperature
then remains relatively stable for the rest of the evolution, with oscillations
of order 10\%. The highest spin configuration s.9, however, reaches
significantly higher temperatures, with $T \sim 5MeV$.  
Fig.~\ref{fig:profjands} shows the entropy and specific angular momentum
profile of three of our disks at the end of the simulation.
At the final time, each of the
disks shows an inverted entropy gradient in the inner region between the
black hole and the radius of maximum surface density.  The entropy difference
between the inner edge of the disk and the density maximum is about 10\%.
However, the disk is at least 
partially stabilized by the strong shear in the rotational velocity in
these inner regions.

\begin{figure}
\includegraphics[width=7.3cm]{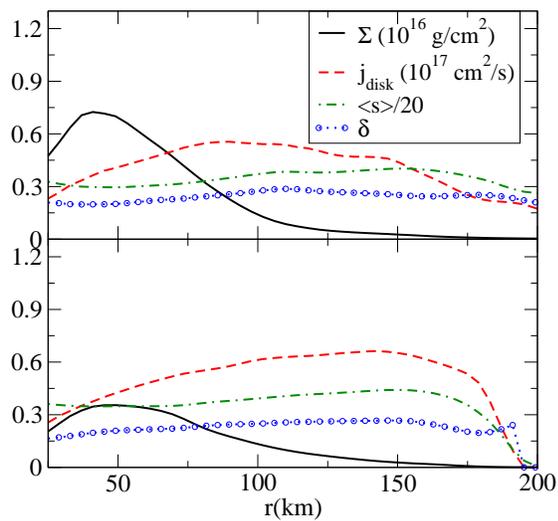}
\caption{
  Profile of the disk forming in the simulation s.5i60 at $t=20{\rm ms}$
(upper panel) and $t=40{\rm ms}$ (lower panel).
}
\label{fig:profcomp}
\end{figure}

In Fig.~\ref{fig:profcomp}, we show two snapshots of the
disk profile for the s.5i60 simulation at, respectively, 20ms and 40ms after 
merger. As the disk keeps accreting, the surface density decreases but the profile
is otherwise mostly constant. The small inverted entropy gradient and the strong 
positive specific angular momentum gradient of the inner disk are
visible, while outside the radius of maximum density the entropy
profile is mostly constant and the specific angular momentum increases
more slowly until $r\approx 100-150$km.  Beyond this radius, matter is still
in the remnant tidal tail rather than the settled disk.  The time evolution
of these two quantities is extremely small, the only difference being
a smoother profile at late times. The disk is relatively thick, with 
$H/r \sim 0.2$ at all radii within the disk, a value that remains constant from
a few milliseconds after merger to the end of the simulation. 

\begin{figure}
\includegraphics[width=7.3cm]{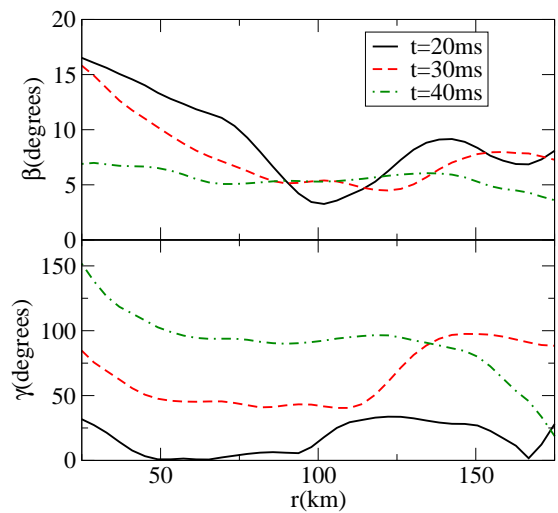}
\caption{
  {\it Upper panel:} Tilt profile of the accretion disk formed in simulation 
s.5i60, as obtained from evolutions on a static background metric. The 
inclination of the disk decreases in time, with $\beta \approx 10-15^\circ$
at the time of disk formation, but $\beta \approx 5-7^\circ$ towards the end 
of the simulation.\\
{\it Lower panel:} Twist profile for the same configuration. Over the 20ms of 
evolution, the disk goes through more than one fourth of a precession period. 
}
\label{fig:tilt}
\end{figure}

For inclined disks, we also measure the tilt $\beta$ and twist $\gamma$ of the
disk, as defined by Eq.~(\ref{eq:jdisk}).
For large inclinations (s.5i60 and s.5i80), the tilt angle between the disk 
angular momentum and the orientation of the BH spin varies slowly during the 
evolution of the disk, at least in the higher density regions.  
Fig.~\ref{fig:tilt} shows the tilt and twist profiles 20, 30 and 40ms 
after merger for simulation s.5i60. In the inner part of the disk 
($r<50{\rm km}$), the tilt decreases from $15^\circ$ to $7^\circ$
over those 20ms.
The precession of the disk is more significant: we observe
a variation of the twist of about $100^\circ$ over the same period. 
However, the precession rate is not constant at all radii and changes in
time; the disk has not reached a state in which it precesses at a constant rate
as one solid body.
For smaller inclinations, relative variations in the tilt
are larger. The inclination of the disk decreases at late-time to
$\beta < 5^\circ$, and no global precession is observed.

An inclined  disk showing some similarities with the 
results of our most inclined simulations (s.5i60 and s.5i80) was evolved in 
the presence of magnetic fields by 
Fragile {\it et al.}~\cite{2009ApJ...691..482F}. Even though the two
simulations vary greatly in their initial conditions --- Fragile {\it et al.}
start their simulation from a torus of matter with peak density at $r=25M$
($\sim200{\rm km}$) --- the thickness and 
inclination of the disks are equivalent and some results
from \cite{2009ApJ...691..482F} could apply to the late time behavior of our
disks. In Fragile {\it et al.}, the inner disk is
warped by the gravitomagnetic torque of the black hole, leading to larger 
tilts at lower radii. The same torque leads to a precession of the disk over a 
period of about 4s (for a black hole remnant of final mass 
$M_{\rm BH}=5.6M_\odot$). This last effect is, 
however, acting over time scales longer than the lifetime of the disk formed in 
BHNS mergers, so that it seems unlikely that our disk would have time to reach 
the steady precession described in \cite{2009ApJ...691..482F}. The 
magneto-rotational instability (MRI), on the other hand, appears to develop 
over roughly one orbital time scale, or about $20{\rm ms}$ for the initial 
configuration chosen in \cite{2009ApJ...691..482F} and a central black hole
of mass $M_{\rm BH}=5.6M_\odot$. The rise of the MRI might be even faster
for a disk more similar to the results of our simulations, as the peak of the 
density profile is significantly closer to the black hole in our disks than 
in \cite{2009ApJ...691..482F}, and the evolution time scale is thus shorter: the
orbital period of circular orbits at $r=50{\rm km}$ is about 5ms. 
The MRI should have a strong influence on the redistribution of angular 
momentum in the fluid, and therefore on the accretion rate.  However, the
accretion rate is also influenced by the presence of
shocks in the disk (see e.g. Fragile and Blaes~\cite{2008ApJ...687..757F} for 
shocks in tilted
disks similar to \cite{2009ApJ...691..482F}). This means that interactions
between the disk and the matter falling back in the tidal tail are also likely 
to play an important role in the determination of the lifetime of the disk.
Thus, both the magnetic effects and realistic initial conditions are required
to accurately predict the lifetime of the disks resulting from BHNS mergers.

\section{Conclusions}
\label{conclusions}


Astrophysical BHNS binaries are expected to have BH spins that
are not aligned with the orbital angular momentum of the system.
We performed here the first fully general relativistic simulations
of BHNS systems with precessing orbits. We find that for
realistic inclinations of the BH spin with respect to the initial
orbital angular momentum ($\phi_{\rm BH}=0-80^\circ$), a mass ratio 
of 1:3, and a moderate black hole spin $a_{\rm BH}/M_{\rm BH}=0.5$, the mass 
of the disk varies by about a factor of 2. More important, the inclination
of the spin seems to have a significant impact on the final
remnant only for $\phi_{\rm BH}>40^\circ$. According to population
synthesis models by Belczynski {\it et al.}~\cite{2008ApJ...682..474B},
this means that, for binaries with initial spin of 0.5, half of
the systems would have disks nearly as massive as if the BH spin was aligned
with the orbital angular momentum of the system.  This confirms the relevance
of the aligned BHNS studies previously undertaken by ourselves and by
other groups.  At late times, the 
inclination of the disks formed in these precessing systems remain
relatively modest ($\beta < 15^\circ$). Most
of the angular momentum of the system is in the orbital motion of the binary,
which precesses around the total angular momentum of the system at a small
misalignment angle ($\sim 5-20^\circ$ for $\phi_{\rm BH}=20-80^\circ$).
The black hole spin axis itself is inclined at a larger angle to
the total angular momentum, 
but its misalignment decreases as the hole accretes matter from the disrupted 
neutron star. This suggests that more inclined disks could be observed for
larger black hole spins or more extreme mass ratios. Here, our most inclined
binaries have an average tilt $\beta\sim 10^\circ$. 
From the results of 
Fragile {\it et al.}~\cite{2009ApJ...691..482F},
we would expect those disks to precess as one solid body around the black hole
--- but only over time scales far longer than the expected lifetime of our
post-merger disks
($\tau_{\rm prec} \sim 4s >> \tau_{\rm acc}\sim 100{\rm ms}$).

For spins aligned with the orbital angular momentum, our study shows
qualitative agreement with previous results by 
Etienne {\it et al.}~\cite{Etienne:2008re}. As expected, large spins favor
the formation of a massive disk. By studying a higher initial
spin ($a_{\rm BH}/M_{\rm BH}=0.9$), we also show that large disks of mass
$M_{\rm disk}\approx 0.5M_\odot$ can be obtained for $M_{\rm BH}=4.2M_\odot$.

All the cases studied here produce post-merger systems that are promising
SGRB central engines.  Despite large differences 
in the disk masses, the lifetime of the system seems mostly
independent of the black hole spin; we find an accretion time scale
$\tau_{\rm acc}\sim 75-150{\rm ms}$ for all cases. The disks have a 
peak surface density located about $50{\rm km}$ away from the hole, and extend about twice
as far. They are always thick ($H/r\approx 0.2$), hot 
($\langle T \rangle=3-5$MeV), and
accreting at a super-Eddington rate ($\dot M=0.5-5M_\odot/s$). 
All simulations also show the presence of a baryon-free
region, at least at densities above the threshold at which atmospheric 
corrections begin to have an impact ($\sim 10^9{\rm g}/{\rm cm}^3$).
This region covers a cone with an opening angle of $30-50^\circ$ around
the axis of the black hole spin, except for the high-spin configuration for which
the disk is significantly closer to the BH, and the opening angle varies within the range $5-10^\circ$. 
Such a region is required if relativistic jets are to
be launched.

Accretion continues throughout the disk evolution, but the
thermal and rotational profiles do seem to stabilize. 
The specific angular momentum of the disks increases with radius,
so they are not subject to the Rayleigh instability.
The angular velocity, however, decreases with radius, so these disks
are subject to the magneto-rotational instability (MRI), an effect
not included in our simulations.

The late-time behavior of the black hole-accretion disk system is critical
if we want to understand the potential of BHNS mergers as progenitors for short
gamma-ray bursts. Currently, the measurement of the properties of the disk and 
their evolution in time suffers from the limitations of our simulations.
The general characteristics of the disk can be obtained, but a more detailed
evolution would certainly require the inclusion of magnetic fields
and neutrino radiation. These effects will be added to our evolutions in the
near future.   

\acknowledgments
We thank Geoffrey Lovelace and Harald Pfeiffer for
useful discussions and suggestions. 
This work was supported in part by a grant from the Sherman Fairchild
Foundation, by NSF Grants Nos. PHY-0652952 and PHY-0652929, and NASA Grant No.
NNX09AF96G. This research
was supported in part by the NSF through
TeraGrid~\cite{teragrid} resources provided by
NCSA's Ranger cluster under Grant No. TG-PHY990007N.
Computations were also performed on the GPC
supercomputer at the SciNet HPC Consortium.  SciNet is funded by: the
Canada Foundation for Innovation under the auspices of Compute Canada;
the Government of Ontario; Ontario Research Fund - Research Excellence;
and the University of Toronto.

\bibliography{References/References}

\end{document}